\documentclass[aps,amssymb,amsmath,prd,twocolumn,
showpacs,preprintnumbers,superscriptaddress,nofootinbib,floatfix]{revtex4-1}
\usepackage{graphicx}
\usepackage{orcidlink} 
\usepackage{color}
\usepackage{times}
\usepackage{amsmath}
\usepackage[normalem]{ulem}
\usepackage{inputenc}
\usepackage{bm}
\usepackage{multirow}
\usepackage{float}
\usepackage{url}
\usepackage{natbib}
\usepackage{tikz}
\usepackage{tikz-3dplot}
\usetikzlibrary{arrows.meta}
\usetikzlibrary{angles}
\usepackage{wrapfig}

\hypersetup{colorlinks=true,linkcolor=red,urlcolor=blue,citecolor=blue}

\newcommand{\ie}{i.e.,~}
\newcommand{\eg}{e.g.,~}
\newcommand{\ep}{\varepsilon}
\newcommand{\nb}{n_{\rm B}}
\newcommand{\mub}{\mu_{\rm B}}
\newcommand{\mb}{m_{\rm B}}
\newcommand{\csq}{c^2_{s}}

\newcommand{\apjl}{Astrophys. J. Lett.}
\newcommand{\apjs}{Astrophys. J., Supp.}
\newcommand{\aap}{Astron. Astrophys.}

\newcommand{\mnras}{Mon. Not. R. Astron. Soc.}
\newcommand{\nphysa}{Nuclear Physics A}
\newcommand{\Msun}{{\rm M}_{\odot}}
\def\fm3{\;\text{fm}^{-3}}

\definecolor{lime}{HTML}{A6CE39}
\DeclareRobustCommand{\orcidicon}{%
	\begin{tikzpicture}
	\draw[lime, fill=lime] (0,0) 
	circle [radius=0.17] 
	node[white] {{\fontfamily{qag}\selectfont \tiny ID}};
	\draw[white, fill=white] (-0.0625,0.095) 
	circle [radius=0.008];
	\end{tikzpicture}
	\hspace{-2mm}
}
\foreach \x in {A, ..., Z}{
	\expandafter\xdef\csname orcid\x\endcsname{\noexpand\href{https://orcid.org/\csname orcidauthor\x\endcsname}{\noexpand\orcidicon}}
}

\begin{document}

\title{Three-dimensional equation of state extension of quark matter in Fermi-liquid theory}

\author{Zhenyu Zhu \orcidlink{0000-0001-9189-860X}}
\email{zhenyu.zhu@rit.edu}
\affiliation{
Center for Computational Relativity and Gravitation, Rochester Institute of Technology, Rochester, NY 14623, USA;\\}
\affiliation{Tsung-Dao Lee Institute, Shanghai Jiao Tong University, Shanghai~201210, China}

\author{Shuai Zha \orcidlink{0000-0001-6773-7830}}
\email{zhashuai@ynao.ac.cn}
\affiliation{
International Centre of Supernovae (ICESUN), Yunnan Key Laboratory of Supernova Research, Yunnan Observatories, Chinese Academy of Sciences (CAS), Kunming 650216, People's Republic of China;\\}

\author{Sophia~Han
\orcidlink{0000-0002-9176-4617}}
\email{sjhan@sjtu.edu.cn}
\affiliation{Tsung-Dao Lee Institute, Shanghai Jiao Tong University, Shanghai~201210, China}
\affiliation{School of Physics and Astronomy, Shanghai Jiao Tong University, Shanghai~200240, China}
\affiliation{State Key Laboratory of Dark Matter Physics, Shanghai Jiao Tong University, Shanghai 201210, China}

\date{\today}

\begin{abstract}
The cold, dense matter equation of state (EoS) determines crucial global properties of neutron stars (NSs), including the mass, radius and tidal deformability. However, a one-dimensional (1D), cold, and $\beta$-equilibrated EoS is insufficient to fully describe the interactions or capture the dynamical processes of dense matter as realized in binary neutron star (BNS) mergers or core-collapse supernovae (CCSNe), where thermal and out-of-equilibrium effects play important roles. We develop a method to self-consistently extend a 1D cold and $\beta$-equilibrated EoS of quark matter to a full three-dimensional (3D) version, accounting for density, temperature, and electron fraction dependencies, within the framework of Fermi-liquid theory (FLT), incorporating both thermal and out-of-equilibrium contributions. We compare our FLT-extended EoS with the original bag model and find that our approach successfully reproduces the contributions of thermal and compositional dependencies of the 3D EoS. Furthermore, we construct a 3D EoS with a first-order phase transition (PT) by matching our 3D FLT-extended quark matter EoS to the hadronic DD2 EoS under Maxwell construction, and test it through the GRHD simulations of the TOV-star and CCSN explosion. Both simulations produce consistent results with previous studies, demonstrating the effectiveness and robustness of our 3D EoS construction with PT.
\end{abstract}

\maketitle


\section{Introduction}
Our understanding of strong interactions 
between fundamental particles in the high-density, low-temperature region 
of the Quantum Chromodynamics (QCD) phase diagram 
remains limited due to their inherent complexity. 
Recent multi-messenger observations of the binary neutron star (BNS) merger event GW170817/AT2017gfo~\citep{2017PhRvL.119p1101A, 2017ApJ...851L..21V} have provided a new way of 
constraining dense matter equation of state (EoS) 
through tidal deformability measurements. Furthermore, the NICER mission~\citep{2019ApJ...887L..24M, 2019ApJ...887L..21R, 2021ApJ...918L..28M, 2021ApJ...918L..27R, 2024ApJ...971L..20C} 
for the first time enabled
accurate simultaneous measurements of the NS masses and radii. 
Moreover, recent advancements in QCD theory and terrestrial experiments have also yielded valuable insights~\citep{2024LRR....27....3K, 2025PhRvD.111j3037P, 2022PhRvL.128t2701K}.
Taken together, these advancements have significantly improved our ability to constrain the cold, $\beta$-equilibrated 
EoS of NS matter.

However, dynamical evolution of e.g. proto-neutron stars, core-collapse supernovae (CCSNe)~\citep{1979NuPhA.324..487B,2005ApJ...629..922S,2009A&A...496..475M,2013ApJ...764...99S,2013ApJ...765...29C,2019PhRvC.100e5802S,2020PhRvL.124i2701Y,2020ApJ...894....4D,2025ApJ...980...53E, 2024arXiv241210046C}, and binary neutron star (BNS) mergers~\citep{2016MNRAS.460.3255R, 2017PhRvD..96l4005B, 2018PhRvD..98j4028P, 2021PhRvD.104h3004Z, 2022PhRvD.105j4028S, 2023PhRvD.108f3032B, 2023ApJ...952L..36F, 2023PhRvL.131a1401K, 2024arXiv240614669B, 2024arXiv240918185B, 2024PhRvD.109f4061N, 2025PhRvD.111d3013P, 2025ApJS..276...35F, 2025PhRvD.111j3043J} also sensitively depends on the state of matter at finite temperatures and conditions deviating from neutrinoless $\beta$-equilibrium. During these processes, shock waves form and heat the matter, 
rendering thermal contributions important in certain regions. 
Meanwhile, high temperatures can trigger neutrino emissions, whose rates rely on the chemical potentials of each component. This, in turn, impacts the electron fraction and further contributes to the pressure of matter~\citep{2003MNRAS.342..673R, 2016MNRAS.460.3255R, 2022MNRAS.512.1499R, 2025PhRvD.111d4074C, 2025PhRvL.134g1402C, 2025ApJ...985L..36N}. Such information is not included in the cold $\beta$-equilibrated EoS. To accurately model these processes, a three-dimensional (3D) EoS table that depends on density, temperature, and electron fraction is an essential input for comprehensive hydrodynamic simulations of CCSNe and BNS mergers.

Different nuclear interactions can produce the same cold $\beta$-equilibrium EOS, as extra degrees of freedom become degenerate in this context. To extend the cold $\beta$-equilibrated EoS to 3D form for use in simulations, additional information about the interactions has to be specified. For instance, in Ref.~\citep{2019ApJ...875...12R}, the symmetry energy and Dirac effective mass were parameterized by fitting 9 EoS tables, contributing to deviations from $\beta$-equilibrium and the thermal component of the EoS, respectively. Although these effects do not impact the tidal deformability or radius of NSs, they do play an important role in BNS mergers and CCSNe~\citep{2023PhRvD.108f3032B, 2023ApJ...952L..36F, 2024PhRvD.110d3002R, 2024arXiv240711143G, 2018PhRvL.120d1101A, 2022MNRAS.509.1096M, 2024ApJ...967L..14M, 2025PhRvD.111d4074C, 2025PhRvL.134g1402C}, and might be observable through the post-merger gravitational waves (GWs).

It has been proposed that quark matter might 
exist in 
the interior of compact stars, 
for instance as a dense quark core within a hybrid star~\citep{2019PhRvL.122f1101M, 2019PhRvD..99h3014H, 2019PhRvD..99j3009M, 2019PhRvL.122f1102B, 2020NatPh..16..907A, 2020PhRvD.101j3006E, 2020ApJ...904..103M, 2020ApJ...899..164H, 2021PhRvD.103f3026T, 2021PhRvD.104h3029P, 2022PhRvL.129r1101H}, or as a self-bound strange quark star~\citep{2009PhRvL.103a1101B, 2010PhRvD..81b4012B, 2017ApJ...846..163W, 2018ApJ...852L..32D, 2018RAA....18...24L, 2019ApJ...881..122D, 2021PhRvD.103l3011Z, 2021PhRvD.104h3004Z, 2022PhRvD.106j3032B, 2022PhRvD.106j3030Z, 2023PhRvD.108f4007C, 2024arXiv240711143G}. 
The astrophysical implications of quark matter in compact stars have been extensively discussed in various contexts, including inspiral GW signal~\citep{2020ApJ...904..103M, 2019PhRvD..99h3014H, 2023PhRvD.107b4025U}, mass-radius measurements~\citep{2021arXiv210813071J, 2021PhRvD.104f3003L}, post-merger GW emission~\citep{2019PhRvL.122f1101M, 2019PhRvL.122f1102B, 2020PhRvL.124q1103W}, et cetera. Compared to hadronic matter, quark matter 
is supposed to emerge 
only at 
sufficiently high density, where asymptotic freedom and weak interactions between quarks simplify its description and parameterization. Although there are numerous models and parametrizations that can describe quark matter with large uncertainties, most of them have focused on the zero-temperature and $\beta$-equilibrated EoS.
These finite temperature and out-of-equilibrium effects were less explored, despite their significance for BNS mergers and CCSNe, as discussed in several studies~\citep{2023PhRvD.108f3032B, 2023ApJ...952L..36F, 2024PhRvD.110d3002R, 2024arXiv240711143G}.

Landau Fermi-liquid theory (FLT) has been considerably successful in describing interacting fermionic many-body systems~\citep{landau1980course,baym2008landau}. In FLT, the interaction between a fermion and its surroundings makes it behave like a dressed particle surrounded by a cloud of other particles. These fermions, enveloped by their interaction clouds, are considered to be quasiparticles with modified masses and other dynamical properties. The low-lying excited states at low temperatures (when the temperature $T$ is much smaller than the chemical potential $\mu$) can be effectively described in terms of quasiparticle excitations. These characteristics make FLT suitable for describing dense nuclear matter and quark matter in neutron stars~\citep{1976NuPhA.262..527B, 2012PhRvD..85l5030V, 2015PhRvC..92b5801C, 2017IJMPE..2640005C, 2019PhRvC.100f5807F}
(note though non-Fermi liquid effects may arise under some certain conditions in dense QCD~\cite{2007NuPhA.785..110S}),
as in such environments, effective nucleons (under the mean-field approximation) and quarks interact weakly, and temperatures, even in the remnants of BNS mergers and CCSNe, remain not as high as comparable to the chemical potential.

This paper is organized as follows. 
We describe the theoretical framework of Fermi-liquid theory for multi-component system and its application to the extension of quark matter EoS to finite temperature in Sec.~\ref{sec:FLT}. 
We then compare the FLT-extended EoSs with the original bag model ones in Sec.~\ref{sec:comp}. 
Section~\ref{sec:PT} presents the construction of a full 3D EoS with phase transition (PT) and discusses its properties. 
In Sec.~\ref{sec:simulation}, we apply 
the constructed 3D EoS with PT
to GRHD 
simulations of a TOV star as well as CCSN simulations. 
Finally, we summarize and conclude in Sec.~\ref{sec:cons}.

Before proceeding to the next section, we 
clarify the notations for different types of energy used throughout this paper to avoid potential confusion. Specifically, three types of energy appear as: the single-particle energy, denoted by $e$; the energy density of dense matter, denoted by $\ep$; and the specific internal energy, denoted by $\epsilon$.

\section{Multi-component Fermi-liquid theory}
\label{sec:FLT}

The current mass of quarks is significantly smaller than its kinetic energy, making quark matter a relativistic system. Furthermore, both the electron fraction $Y_e$ and the baryon number density $\nb$ are treated as independent variables in the 3D EoS tables, necessitating the consideration of at least two components in quark matter. Accordingly, we begin directly with the multi-component relativistic Fermi-liquid theory. 
A similar introduction can also be found in Refs.~\citep{2012PhRvD..85l5030V, 2019PhRvC.100f5807F}. In this paper, we adopt the units where $\hbar=c=k_B=1$. 

\subsection{The Landau parameters}
The Fermi liquid theory (FLT) starts from an energy functional
\begin{eqnarray}
  \label{eq:e_functional}
  \delta E(\bar{n}_{i}) = \sum_{\vec{p}_i,i} e_i^{(0)}(\vec{p}_i) \delta \bar{n}_i + \frac{1}{2} \sum_{\vec{p}_i,\vec{p}_{i'},i,i'} f_{i,i'}(\vec{p}_i,\vec{p}_{i'}) \delta \bar{n}_i \delta \bar{n}_{i'},
\end{eqnarray}
where $i=u, d, s, e$ denotes each particle component; up, down, strange quarks and electrons are included in this case. The function $\bar{n}_i = 1/(e^{(e_i - \mu_i)/T}+1)$ represents the Fermi distribution function, and $\delta \bar{n}_i$ is its variation. Note that Eq.~(\ref{eq:e_functional}) is a straightforward expansion of the energy with respect to the distribution $\bar{n}_i$ near the ground state. The first order coefficient $e_i^{(0)}=\delta E/\delta \bar{n}_i$ hence represents the chemical potential $\mu_i$, and the second order coefficient $f_{i,i'}$ represents the interaction between the excited particles.

Consider the ground state to be a system with zero temperature and number density $n_i$ for each component. 
As we add a few more particles, 
the distribution functions are varied as
\begin{eqnarray}
  \label{eq:df_ni}
  \delta \bar{n}_i =
  \begin{cases}
    1, \qquad {\rm when}\ e_i(p_{F_i}) < e_i(p) < e_i(p_{F_i}) + \delta e ; \\
    0, \qquad {\rm else}.
  \end{cases}
\end{eqnarray}
In FLT, the interaction was turned on adiabatically. Hence, there is a one-to-one mapping between the non-interacting system (Fermi gas) and the interacting system (Fermi liquid). This mapping 
guarantees the relation between the Fermi momentum and the number density $n_i=g p_{F_i}^3/6\pi^2$, where $g$ is the degree of degeneracy. Thereafter, the conditions of $\delta \bar{n}_i$ being one can be rewritten as $p_{F_i} < p < p_{F_i} + \delta p_{F_i}$. We plug this $\delta \bar{n}_i$ into Eq.~(\ref{eq:e_functional}) and rewrite it as
\begin{eqnarray}
  \label{eq:e_df_ni}
  \delta E(\bar{n}_{i}) & = & \frac{g}{2\pi^2} \sum_i \int_{\vec{p}_{F_i}}^{\vec{p}_{F_i} + \delta \vec{p}_{F_i}} e_i^{(0)}(p_i) p_i^2 dp_i \nonumber \\
  & & + \frac{1}{2} \frac{g^2}{(8\pi^3)^2} \sum_{ii'} \int_{p_{F_i}}^{p_{F_i} + \delta p_{F_i}} \times \nonumber \\
  & & \int_{\vec{p}_{F_{i'}}}^{\vec{p}_{F_{i'}} + \delta \vec{p}_{F_{i'}}} f_{i,i'}(p_i, p_{i'}) d^3p_i d^3p_{i'}.
\end{eqnarray}
In the first term of Eq.~(\ref{eq:e_df_ni}), the coefficient $e_i^{(0)}(p_i)$ can be expanded near the Fermi surface
\begin{eqnarray}
  \label{eq:exp_eps}
  e_i^{(0)}(p_i) = e_{F_i} + \frac{p_{F_i}}{m_i^\ast}(p_i-p_{F_i}),
\end{eqnarray}
where $e_{F_i}=e_i^{(0)}(p_{F_i})$, and $m_i^\ast$ is the Landau effective mass. We then expand the first term as 
\begin{eqnarray}
  \label{eq:e_df_ni_i}
  & & \frac{g}{2\pi^2} \sum_i \int_{\vec{p}_{F_i}}^{\vec{p}_{F_i} + \delta \vec{p}_{F_i}} e_i^{(0)}(p_i) p_i^2 dp_i \nonumber \\
  & = & \sum_i \frac{g}{2\pi^2} \left[e_{F_i} \int p_i^2dp_i + \frac{p_{F_i}}{m_i^\ast} \int(p_i-p_{F_i})p_i^2dp_i  \right] \nonumber \\
  & = & \sum_i \frac{g}{2\pi^2} \left[e_{F_i} p_{F_i}^2 \delta p_{F_i} + \left(e_{F_i} + \frac{p_{F_i}^2}{2m_i^\ast} \right) p_{F_i} \delta p_{F_i}^2 \right].
\end{eqnarray}
Note that we only keep up to $\mathcal{O}(\delta p_{F_i}^2)$ order in the last line of Eq.~(\ref{eq:e_df_ni_i}).

In the second term of Eq.~(\ref{eq:e_df_ni}), the coefficient $f_{i,i'}(\vec{p}_i, \vec{p}_{i'})$ is angular dependent and we expand it near the Fermi surface to the first order of $\cos\theta$
\begin{eqnarray}
  \label{eq:f_expand}
f_{i,i'}(\vec{p}_{F_i}, \vec{p}_{F_{i'}}) & = & \sum_l f_{l, ii'}(p_{F_i},p_{F_{i'}})P_l(\cos\theta) \nonumber \\
& \approx & f_{0, ii'}(p_{F_i},p_{F_{i'}}) + f_{1, ii'}(p_{F_i},p_{F_{i'}})\cos\theta, \nonumber \\
\end{eqnarray}
where $\theta$ is the angle between $\vec{p}_{F_i}$ and $\vec{p}_{F_{i'}}$ and $P_l(\cos\theta)$ represents the Legendre polynomials. Plug Eq.~(\ref{eq:f_expand}) into the second term, and it yields
\begin{eqnarray}
  \label{eq:e_df_ni_ii}
  & & \frac{1}{2} \frac{g^2}{(8\pi^3)^2} \sum_{ii'} \int \int f_{i,i'}(p_i, p_{i'}) d^3p_i d^3p_{i'} \nonumber \\ & = & \sum_{ii'} \frac{g^2}{16\pi^4} p_{F_i}^2  p^2_{F_{i'}} \delta p_{F_i} \delta p_{F_{i'}} \int f_{i,i'}(\vec{p}_{F_i}, \vec{p}_{F_{i'}}) \sin\theta  d\theta \nonumber \\
           & = & \sum_{ii'} \frac{1}{2} \frac{g^2}{4\pi^4} f_{0,ii'}(\vec{p}_{F_i}, \vec{p}_{F_{i'}}) p_{F_i}^2  p^2_{F_{i'}} \delta p_{F_i} \delta p_{F_{i'}}.
\end{eqnarray}
Accordingly, we only keep terms
up to $\mathcal{O}(\delta p_{F_i}^2)$ 
in the expression. 
By combining Eq.~(\ref{eq:e_df_ni_i}) and Eq.~(\ref{eq:e_df_ni_ii}), the expression for $\delta E$ is simplified as
\begin{eqnarray}
  \label{eq:de_taylor}
  \delta E = \sum_i \frac{dE}{dp_{F_i}} \delta p_{F_i} + \frac{1}{2} \sum_{ii'} \frac{d^2E}{dp_{F_i} dp_{F_{i'}}} \delta p_{F_i} \delta p_{F_{i'}},
\end{eqnarray}
where
\begin{eqnarray}
  \label{eq:dedp}
  \frac{dE}{dp_{F_i}} & = & \frac{g}{2\pi^2} e_{F_i} p_{F_i}^2, \\
  \label{eq:d2edp2}
  \frac{d^2E}{dp_{F_i} dp_{F_{i'}}} & = & \frac{g}{2\pi^2}\left( 2e_{F_i} + \frac{p_{F_i}^2}{m_i^\ast}\right) p_{F_i} \delta_{ii'} \nonumber \\
  & & + \frac{g^2}{4\pi^4} f_{0,ii'}(p_{F_i}, p_{F_{i'}}) p_{F_i}^2  p^2_{F_{i'}}.
\end{eqnarray}
Furthermore, we can calculate the chemical potential
\begin{eqnarray}
  \label{eq:mu}
  \mu_i = \frac{dE}{dn_i} = \frac{2\pi^2}{g} \frac{1}{p_{F_i}^2} \frac{dE}{dn_i} = e_{F_i},
\end{eqnarray}
and the derivative of the chemical potential
\begin{eqnarray}
  \label{eq:dmudp}
  \frac{d\mu_i}{dp_{F_{i'}}} & = & \frac{p_{F_i}}{m_i^\ast} \delta_{ii'} + \frac{g}{2\pi^2} f_{0,ii'} p_{F_{i'}}^2,
\end{eqnarray}
or
\begin{eqnarray}
  \label{eq:dmudn}
  \frac{d\mu_i}{dn_{i'}} & = & \frac{2\pi^2}{g} \frac{1}{p_{F_i} m_i^\ast} \delta_{ii'} + f_{0,ii'}(p_{F_i}, p_{F_{i'}}).
\end{eqnarray}
It is noteworthy that the first term refers to the contribution from the Fermi gas, while the second term represents the interactions. 

The Landau effective mass $m_i^\ast$ is defined as
\begin{eqnarray}
  \label{eq:def_mstar}
  m_i^\ast & = & \left(\frac{d e_i^{(0)}(p_i)}{dp_i}\biggr|_{p_i=p_{F_i}} \right)^{-1} p_{F_i}.
\end{eqnarray}
Its connection with the interaction function $f_{i,i'}(\vec{p}_{F_i}, \vec{p}_{F_{i'}})$ can be obtained by applying a Lorentz boost on the system. For more detailed derivations, see Appendix~\ref{app:A}. The final expression for $m_i^\ast$ and $f_{1,ii'}$ can be written as
\begin{eqnarray}
  \label{eq:emass}
  p_{F_i} & = & \sum_{i'} \frac{g}{6\pi^2} \mu_{i'} p_{F_{i'}}^2 f_{1,ii'} + \mu_i \frac{p_{F_i}}{m^\ast_i}.
\end{eqnarray}
The functions $f_{1,ii'}(p_{F_i}, p_{F_{i'}})$ and $f_{0,ii'}(p_{F_i}, p_{F_{i'}})$ are multi-component versions of the Landau parameters.  
Eq.~(\ref{eq:emass}) can also be written in a form analogous to Eq.~(25) of Ref.~\citep{2019PhRvC.100f5807F}
\begin{eqnarray}
  \label{eq:emass2}
  m^\ast_i & = & \mu_i + \sum_{i'} g \frac{f_{1,ii'}(p_{F_i}, p_{F_{i'}})}{6\pi^2} \frac{\mu_{i'} p_{F_{i'}}^2 m_i^\ast}{p_{F_i}}.
\end{eqnarray}
\subsection{Thermal excitations}
In FLT, the interactions could shift the energy level $e_i^{(0)}(\vec{p}_i)$, resulting 
in its density dependency. 
Despite this modification, the one-to-one mapping between the Fermi gas and the Fermi liquid preserves the forms of function for various thermodynamic quantities. For instance, the entropy density expressed in terms of the distribution function $\bar{n}_i$ is given by
\begin{eqnarray}
  \label{eq:ent}
  s_{v} & = & -\sum_i \int_0^\infty D_i(e_i)\biggl[\bar{n}_i \ln(\bar{n}_i) \nonumber \\
  & & \quad\quad\quad\quad\quad\quad + (1-\bar{n}_i)\ln(1-\bar{n}_i)\biggr]de_i,
\end{eqnarray}
where $D_i(e_i)$ denotes the density of state and $D_i(e_i) = g\,m_i^\ast \,p_{F_i}/(2\pi^2)$ on the Fermi surface. Consider the variation of entropy due to thermal excitations, we have
\begin{eqnarray}
  \label{eq:dent}
  \delta s_v = \sum_i \int_0^\infty D_i(e_i) \frac{e_i - \mu_i}{T} de_i \delta \bar{n}_i ,
\end{eqnarray}
where
\begin{eqnarray}
  \label{eq:dnbar}
  \delta \bar{n}_i = \frac{\partial \bar{n}_i}{\partial e_i} \left[-\frac{e_i-\mu_i}{T} \delta T + \delta e_i - \delta \mu_i \right].
\end{eqnarray}
Given that the last two terms ($\delta e_i - \delta \mu_i$) only contribute at the $\mathcal{O}(T^2)$ or higher order, 
they will be neglected in the following. 
Hence
\begin{eqnarray}
  \label{eq:dent2}
  \delta s_v & = & -\sum_i \int_0^\infty D_i(e_i) \left(\frac{e_i - \mu_i}{T}\right)^2 \frac{\partial \bar{n}_i}{\partial e_i} de_i \delta T  \nonumber \\
  & \approx & \sum_i \left[D_i(\mu_i) \int_{-\infty}^\infty \xi_i \frac{\partial \bar{n}_i}{\partial \xi_i} d\xi_i + \mathcal{O}(T^2) \right] \delta T, \nonumber \\
  & = & \sum_i \left[ \frac{\pi^2}{3} D_i(\mu_i) + \mathcal{O}(T^2) \right] \delta T,
\end{eqnarray}
and
\begin{eqnarray}
  \label{eq:entropy}
   s_v & = & \frac{gT}{6} \sum_i m_i^\ast \,p_{F_i} + \mathcal{O}(T^3),
\end{eqnarray}
where in Eq.~(\ref{eq:dent2}) $\xi_i=(e_i-\mu_i)/T$. 

Other quantities can be derived from 
the entropy through the Maxwell relations. We fix the number density $n_i$ and yield the expression of thermal energy density $\ep_{\rm th}$ as
\begin{eqnarray}
  \label{eq:energy_th}
  \ep_{\rm th} & = & \ep - \ep_{\rm cold} = \int_0^{s_v} T ds_{v}' = \frac{g T^2}{12} \sum_i m_i^\ast \,p_{F_i},
\end{eqnarray}
where $\ep_{\rm cold}=\ep(T=0)$. 
Thermodynamic variables such as chemical potential $\mu_i$ and pressure $p$ can then be derived from $\ep$
\begin{eqnarray}
\label{eq:mu_th}
\mu_i^{\rm th} & = & \left(\frac{\partial \ep}{\partial n_i}\right)_{s_v} - e_{F_i}\nonumber \\
   & = &  - T^2 \left(\frac{\pi^2 m_i^\ast}{6p_{F_i}^2} + \sum_{i'} \frac{g\, p_{F_{i'}}}{12} \frac{\partial m_{i'}^\ast}{\partial n_i} \right) , \\
\label{eq:p_th}
p_{\rm th} & = & \ep_{\rm th}-Ts_v-\sum_i \mu_i^{\rm th} n_i \nonumber \\
  & = & T^2 \left[\frac{g}{18} \sum_i m_i^\ast \,p_{F_i} - \frac{g}{12}\sum_{ii'} p_{F_{i'}} n_i \frac{\partial m_{i'}^\ast}{\partial n_i} \right].
\end{eqnarray}

Having determined the SPEs, chemical potentials, and Landau effective masses, 
in principle one can also proceed to 
explore susceptibilities of the constituents, or compute transport coefficients such as the bulk viscosity~\citep{2004PhRvC..69b5802V, 2024PhRvL.133g1901C} which has been 
revealed to potentially play an important role in BNS mergers~\citep{2025PhRvD.111d4074C, 2025PhRvL.134g1402C}. 
Besides, the current FLT scheme that we develop is mostly applicable 
only to scenarios that involve warm dense matter such as in BNS mergers or CCSNe, for which the low-temperature and high-density conditions required by FLT are still met. In contrast, 
it cannot be directly applied to environments as in heavy-ion collisions, where the temperatures are typically high enough to exceed the degeneracy limit.

\subsection{Single-particle energy of quarks}
The Landau parameters 
cannot be determined 
until the single-particle energies (SPEs) $e_i$ of quasiparticles are known. 
Based on relativistic models of interactions, we assume that $e_i$ takes the following form
\begin{eqnarray}
  \label{eq:pe_rel}
  e_p = \sqrt{p^2 + M^2(\nb)} + V(\nb).
\end{eqnarray}
where $M$ and $V$ are functions of the baryonic number density $\nb$. Although in general SPEs may also depend on the momentum $p$~\citep{2012PhLB..711..104C} through the potential term, here we consider a simplified form where only the dependence on $\nb$ remains.
Extending this form to each quark flavor,
\begin{eqnarray}
  \label{eq:pe_quarks}
  e_i(p) & = & \sqrt{p^2 + M_i^2(\nb)} + V_i(\nb),
\end{eqnarray}
where $i$'s represent $u$, $d$ and $s$ quarks. The functions $M_i$ and $V_i$ may depend on the density of each quark component. However, since the interactions between quarks can be considered to be equivalent across different flavors, their dependence on the density of each flavor should be identical. We therefore apply $M_i=M_i(\nb)$ and $V_i=V_i(\nb)$. Considering the similarity of quarks, 
this relation can be further simplified 
by approximating
\begin{eqnarray}
  \label{eq:q_potentials}
  M_i(\nb) = m_i + M_{\rm diff}(\nb), \quad V_u = V_d = V_s = V,
\end{eqnarray}
where $m_i$ represents the mass of each quark for which we choose $m_u=m_d=0$ MeV, $m_s=100$ MeV, and $M_{\rm diff}$ denotes the difference between the bare quark mass and its Dirac effective masses $M_i$. This formula assumes that quark interactions depend only on the mass of each quark (and also $M_{\rm diff}$ takes the same value for all quarks). At high densities that are typical for the quark phase in NSs, the Dirac effective mass saturates and approaches a constant, \ie $M_{i}(\nb) = M_{i}$. 
Based on this SPE form, the Landau effective mass can be derived and written as~\cite{1976NuPhA.262..527B}
\begin{eqnarray}
  \label{eq:m_eff_q}
  m_i^\ast = \sqrt{p_{F_i}^2 + M_i^2}
\end{eqnarray}

The Landau parameters $f_{0,ii'}$ can then be obtained through the form of chemical potentials
\begin{eqnarray}
  \label{eq:landau_params}
  f_{0,ii'}(p_{F_i}, p_{F_{i'}}) & = & \left(\frac{M_i}{m_i^\ast} M_i' + V' \right) \frac{d\nb}{dn_{i'}} \nonumber \\
  & = & \frac{1}{3} \left(\frac{M_i}{m_i^\ast} M_i' + V' \right),
\end{eqnarray}
where $M_i'=dM_i(\nb)/d\nb$ and $V'= dV(\nb)/d\nb$. However, the other Laudau parameters $f_{1,ii'}$ cannot be determined solely from Eq.~(\ref{eq:emass2}). Since only $m_i^\ast$'s enter the extension of 1D EoS using FLT, we will not discuss the detailed expressions of Landau parameters in subsequent sections.

\subsection{Extension of the quark matter EoS with FLT}
Given a zero-temperature EoS of quark matter at $\beta$-equilibrium, the pressure is a known function of the energy density, i.e., $p = p(\ep)$. Other quantities, such as the baryon chemical potential $\mub$ and the baryon number density $\nb$, can be determined using the thermodynamic relations $\int d\mub/\mub=\int dp/(\ep+p)$ and $\nb=(\ep+p)/\mub$.

The chemical potential of each quark takes the form of Eq.~(\ref{eq:pe_quarks}), where $M_i$'s are constants and predefined, $V(\nb)$ will be determined by the given quark matter EoS. Recall that the conditions for a cold $\beta$-equilibrated EoS are
\begin{eqnarray}
  \label{eq:neutral}
  \frac{2}{3} n_u - \frac{1}{3} n_d - \frac{1}{3} n_s - n_e = 0, \\
  \label{eq:baryon_num}
  \frac{1}{3} n_u + \frac{1}{3} n_d + \frac{1}{3} n_s = \nb, \\
  \label{eq:weak}
  \mu_u + \mu_e = \mu_d; \quad \quad \mu_d = \mu_s.
\end{eqnarray}
These equations describe the conditions for charge neutrality, baryon number conservation, and weak equilibrium. By solving Eqs.~(\ref{eq:pe_quarks}), (\ref{eq:neutral})–(\ref{eq:weak}) together, we can determine the number densities and Fermi momenta of all four particle components (\ie, $u$, $d$, and $s$ quarks, as well as electrons) given the baryon number density $\nb$. The equations can be rewritten as
\begin{eqnarray}
  \label{eq:nu}
  n_u & = & \nb+n_e, \\
  \label{eq:nds}
  n_d+n_s & = & 2\nb-n_e, \\
   \label{eq:muds}
  \sqrt{p_{F_{d}}^2 + M_d^2} & = & \sqrt{p_{F_{s}}^2 + M_s^2}, \\
  \label{eq:muu}
  \sqrt{p_{F_{u}}^2 + M_u^2} + p_{F_{e}} & = & \sqrt{p_{F_{d}}^2 + M_d^2}.
\end{eqnarray}
Solving these four equations can obtain the Fermi momenta $p_{F_u}$, $p_{F_d}$, $p_{F_s}$ and $p_{F_e}$.

The baryonic chemical potential is defined as $\mub=(\partial \ep/\partial \nb)_{s_v}$. Setting the $s_v={\rm const.}$ or equivalently $ds_v=0$, we then have
\begin{eqnarray}
\label{eq:mu_b}
  d\ep & = & \mu_u dn_u + \mu_d dn_d + \mu_s dn_s + \mu_e dn_e \nonumber \\
  & = & \mu_u d\nb + \mu_u dn_e + 2\mu_d d\nb - \mu_d dn_e + \mu_e dn_e \nonumber \\
  & = & (\mu_u + 2\mu_d) d\nb + (\mu_u + \mu_e - \mu_d) dn_e.
\end{eqnarray}
We have applied the conditions of $n_u=\nb+n_e$ and $n_d+n_s=2\nb-n_e$, which are derived from Eqs.(\ref{eq:neutral})--(\ref{eq:baryon_num}). Additionally, the weak equilibrium condition $\mu_d=\mu_s$ is imposed. The timescale of reactions producing $s$ quarks could be as short as $\sim 10^{-6}$s~\citep{2011ApJS..194...39F}, which is much shorter than the hydrodynamic timescale ($\sim$ ms). As a result, this condition is considered to be satisfied at all times, although $\mu_u$ and $\mu_d$ may deviate from weak equilibrium due to the presence of neutrinos. At $\beta$-equilibrium where $\mu_{\beta}=\mu_u + \mu_e - \mu_d=0$, Eq.~(\ref{eq:mu_b}) yields $\mub=\mu_u+2\mu_d$. Then the vector potential can be given after knowing the Fermi momenta
\begin{eqnarray}
  \label{eq:v_eos}
  3V(\nb) = \mub - \sum_{i=u,d,s} \sqrt{p_{F_{i}}^2 + M_i^2}.
\end{eqnarray}

In principle, Eqs.~(\ref{eq:nu})--(\ref{eq:nds}) and (\ref{eq:muds}) can determine the Fermi momenta at any electron fraction $Y_e=n_e/\nb$. Once the vector potential $V(\nb)$ is known, the cold chemical potentials, energy density and pressure at any $Y_e$ can be calculated using
\begin{eqnarray}
\label{eq:mu_cold}
  \mu_i^{\rm cold}(\nb, Y_e) & = & \sqrt{p_{F_{i}}^2 + M_i^2} + V(\nb), \\
\label{eq:e_cold}
  \ep_{\rm cold}(\nb, Y_e) & = & \ep_{\rm cold}(\nb, Y_e^{beta}) + \int_{Y_e^{beta}}^{Y_e} \mu_{\beta} \nb dY', \nonumber \\
  \\
  p_{\rm cold}(\nb, Y_e) & = & \sum_{i=u,d,s,e} \mu_i^{\rm cold} n_i - \ep_{\rm cold}(\nb, Y_e).
\end{eqnarray}
Finally, the thermal contribution for chemical potentials, energy and pressure can be calculated through Eqs.~(\ref{eq:energy_th})--(\ref{eq:p_th}), with the Landau effective mass Eq.~(\ref{eq:m_eff_q}).

\begin{figure*}
\vspace{-0.3cm}
{\centering
\includegraphics[width=0.33\textwidth]{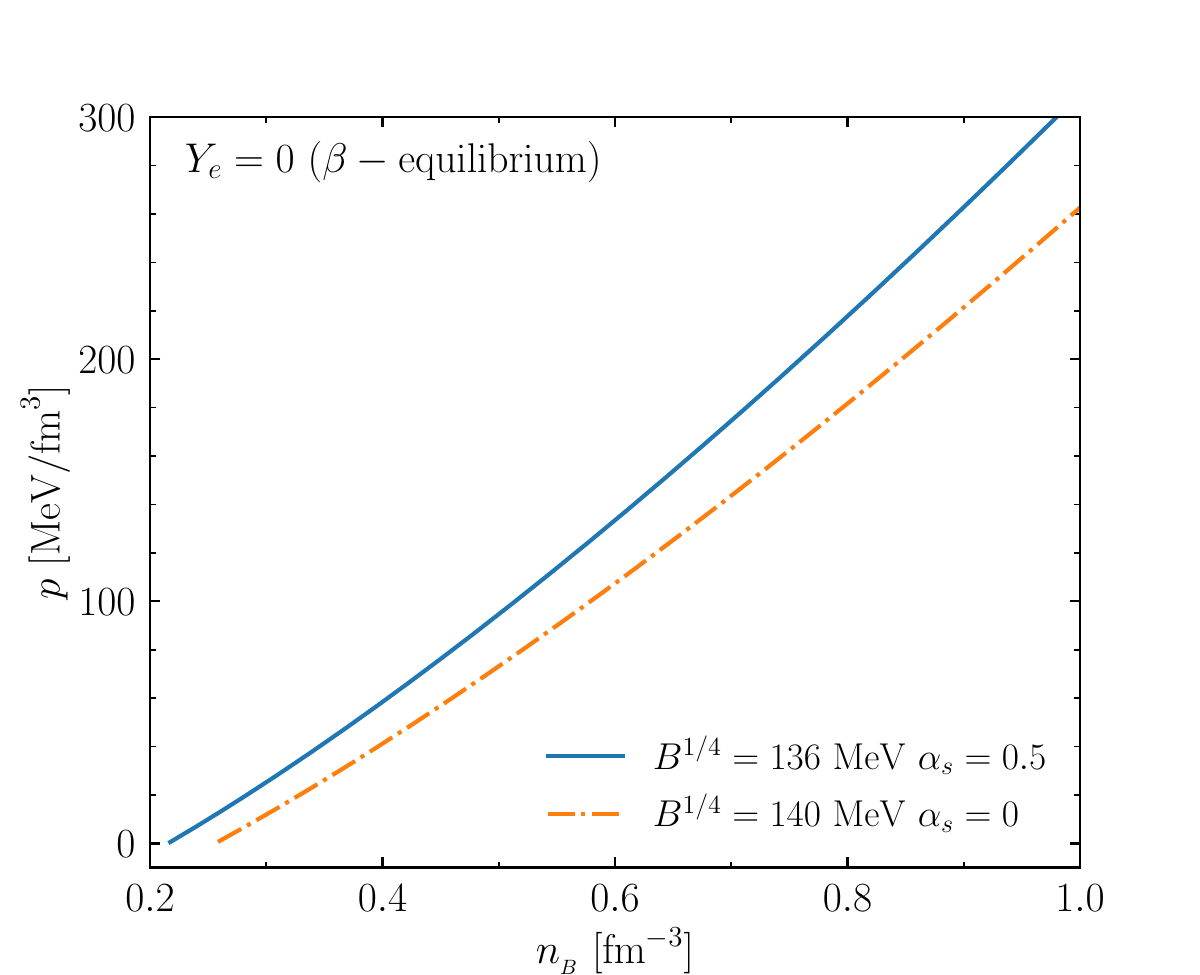}
\includegraphics[width=0.33\textwidth]{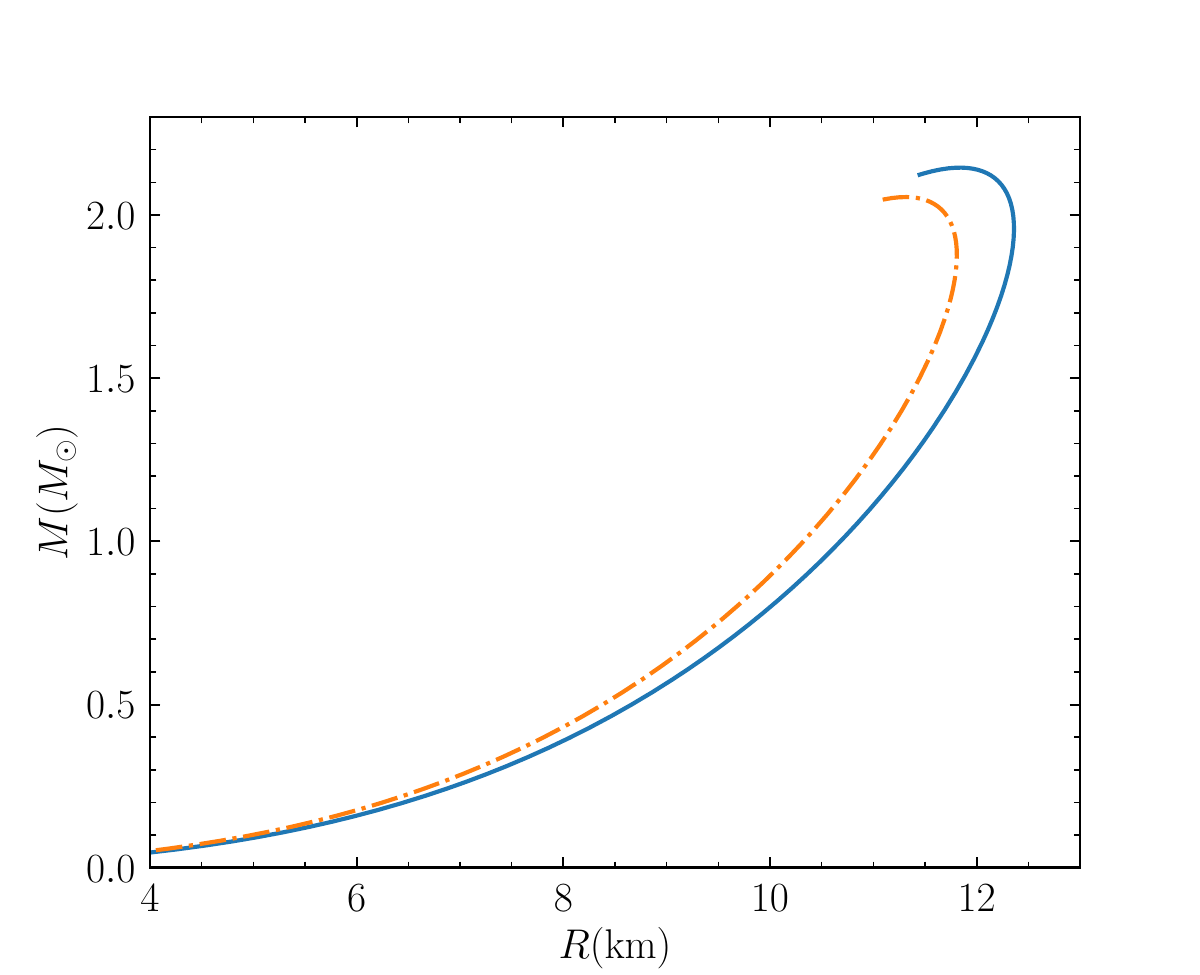}
\includegraphics[width=0.33\textwidth]{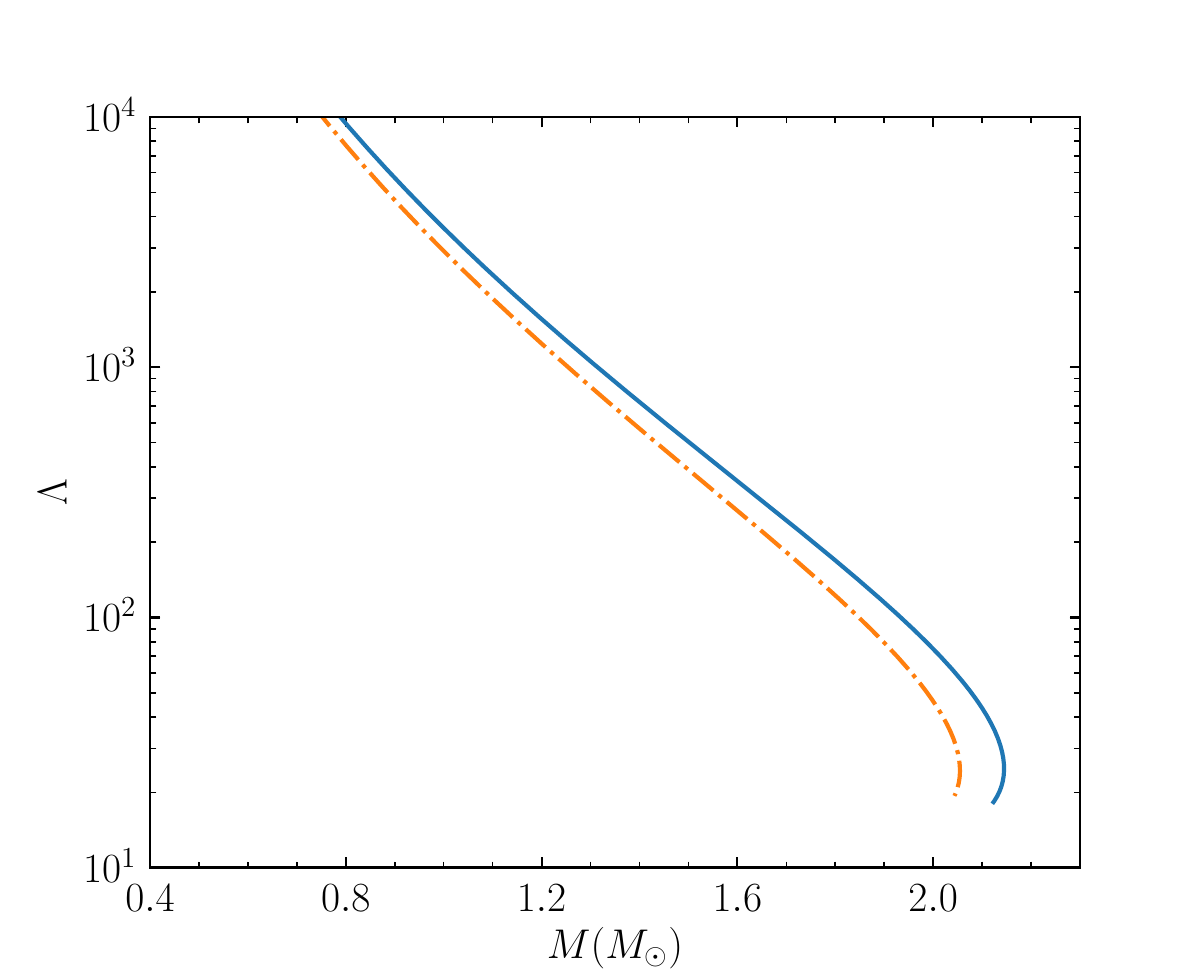}}
\caption{
Pressure as a function of the number density for quark matter EoS using the bag model (left), the mass-radius relation (middle), and the tidal deformability (right) of the corresponding quark stars 
with two parameter sets are displayed.
The blue lines refer to results obtained with $B^{1/4}=136\ {\rm MeV}, \alpha_s=0.5$, and the oranges ones $B^{1/4}=140\ {\rm MeV}, \alpha_s=0$.
}
\label{fig:bag}
\vspace{-0.1cm}
\end{figure*}

\section{Comparison with the bag model}
\label{sec:comp}
In this section, we employ the FLT method to extend the cold $\beta$-equilibrated bag EoS to 3D form and compare 
with the original bag model. 

\subsection{The bag model}
Bag models are phenomenological frameworks used to describe quarks confined within hadrons. Due to their simplicity, bag models have also been widely applied to quark matter in compact star interiors. Because of the asymptotic freedom, quarks move freely inside the bag. However, they are confined and cannot escape to the exterior region. The key idea of the bag model is that the exterior true vacuum and the interior QCD vacuum have an energy difference denoted
by the bag constant $B$.

The bag model can be extended to the finite-temperature case by replacing the step function with the Fermi-Dirac distribution function $\bar{n}_i$
\begin{eqnarray}
\label{eq:e_bag}
  \ep_i(m_i, T, \mu_i) & = & \frac{g}{2\pi^2} \int_0^\infty \sqrt{p^2+m_i^2} p^2 dp \nonumber \\
  & & \times [\bar{n}_i(p,\mu_i) + \bar{n}_i(p,-\mu_i)], \\
  \nonumber \\
  \label{eq:p_bag}
  p_i(m_i, T, \mu_i) & = & \frac{g}{6\pi^2} \int_0^\infty \frac{p^2}{\sqrt{p^2+m_i^2}} p^2 dp \nonumber \\
  & & \times [\bar{n}_i(p,\mu_i) + \bar{n}_i(p,-\mu_i)], \\
  \nonumber \\
  \label{eq:s_bag}
  s_v^i(m_i, T, \mu_i) & = & \frac{g}{2\pi^2} \int_0^\infty p^2 dp \biggl[ -\bar{n}_i(p,\mu_i) \ln \bar{n}_i(p,\mu_i) \nonumber \\
  & & - (1-\bar{n}_i(p,\mu_i))\ln(1-\bar{n}_i(p,\mu_i)) \nonumber \\
  & & - (1-\bar{n}_i(p,-\mu_i))\ln(1-\bar{n}_i(p,-\mu_i)) \nonumber \\
  & & -\bar{n}_i(p,-\mu_i) \ln \bar{n}_i(p,-\mu_i)\biggr], \\
  \nonumber \\
  \label{eq:n_bag}
  n_i(m_i, T, \mu_i) & = & \frac{g}{2\pi^2} \int_0^\infty p^2 dp \nonumber \\
  & & \times [\bar{n}_i(p,\mu_i) + \bar{n}_i(p,-\mu_i)],
\end{eqnarray}

Note that the antiparticle distribution functions $\bar{n}_i(p,-\mu_i)$ are included in the finite-temperature case. The total energy and pressure of quark matter must also include the bag constant,
\begin{gather}
\ep = \sum_i \ep_i(m_i,T,\mu_i) + B, \ p = \sum_i p_i(m_i,T,\mu_i) - B, \\
s_v = \sum_i s_v^i(m_i,T,\mu_i), \quad \nb = \frac{1}{3} \sum_i n_i(m_i,T,\mu_i).
\end{gather}
This simple bag model was improved by including first-order corrections for the strong coupling constant $\alpha_s$~\citep{1984PhRvD..30.2379F, 2011ApJS..194...39F}. The corrected quantities are written as
\begin{eqnarray}
  \label{eq:e_bag_alphas}
  \ep_i(m_i, T, \mu_i, \alpha_s) & = & \ep_i(m_i, T, \mu_i) \nonumber \\
  & & \hspace{-2.5cm} - \biggl[\frac{7}{20} T^4 \pi^2 \frac{50\alpha_s}{21\pi} + \frac{2\alpha_s}{\pi}\biggl(\frac{3}{2} T^2 \mu_i^2 + \frac{3\mu_i^4}{4\pi^2}\biggr) \biggr], \\
  \nonumber \\
  \label{eq:p_bag_alphas}
  p_i(m_i, T, \mu_i, \alpha_s) & = & p_i(m_i, T, \mu_i) \nonumber \\
  & & \hspace{-2.5cm} - \biggl[\frac{7}{60} T^4 \pi^2 \frac{50\alpha_s}{21\pi} + \frac{2\alpha_s}{\pi}\biggl(\frac{1}{2} T^2 \mu_i^2 + \frac{\mu_i^4}{4\pi^2}\biggr) \biggr], \\
  \nonumber \\
  \label{eq:s_bag_alphas}
  s_v^i(m_i, T, \mu_i, \alpha_s) & = & s_v^i(m_i, T, \mu_i) \nonumber \\
  & & \hspace{-0.5cm} - \biggl[\frac{7}{15} T^3 \pi^2 \frac{50\alpha_s}{21\pi} + \frac{2\alpha_s}{\pi} T \mu_i^2 \biggr], \\
  \nonumber \\
  \label{eq:n_bag_alphas}
  n_i(m_i, T, \mu_i, \alpha_s) & = & n_i(m_i, T, \mu_i) \nonumber \\
  & & - \frac{2\alpha_s}{\pi} \biggl(T^2 \mu_i + \frac{\mu_i^3}{\pi^2}\biggr),
\end{eqnarray}
\begin{figure*}
\vspace{-0.3cm}
{\centering
\includegraphics[width=0.48\textwidth]{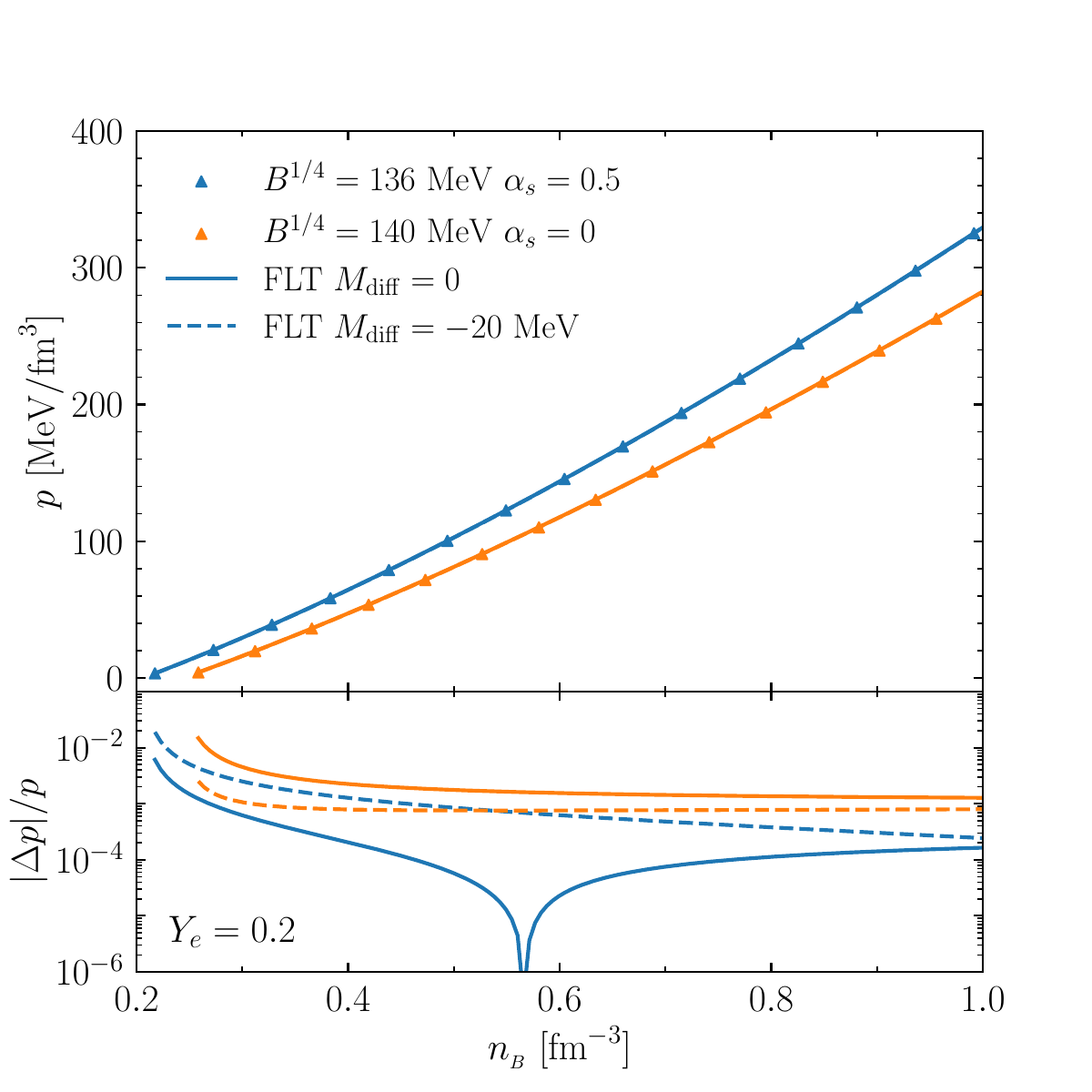}
\includegraphics[width=0.48\textwidth]{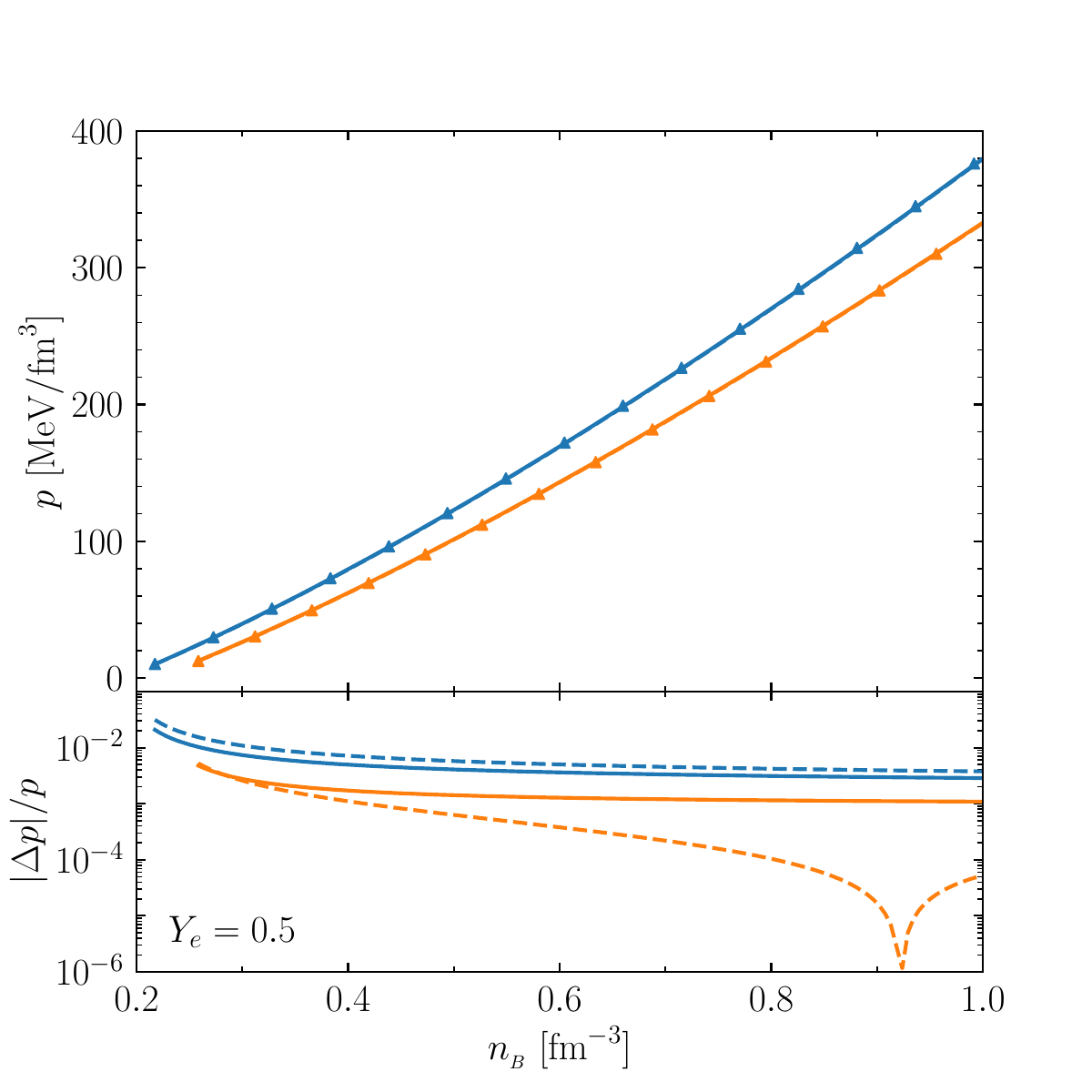}
\caption{Comparison between total pressures in the FLT-extended EoSs (solid and dashed lines) and the original bag model ones (triangular markers) at $Y_e=0.2$ (left) and $Y_e=0.5$ (right); the temperature is held at $T=0.01$ MeV.
Note that results for FLT-extended EoSs with $M_{\rm diff}=0$ MeV (solid) and $M_{\rm diff}=-20$ MeV (dashed) nearly overlap with each other and thus indistinguishable on the upper panels, implying the minor effects of the Dirac effective mass. The lower panels show the deviation of the FLT pressures from those in the corresponding bag model EoSs.}
\label{fig:out_of_equilibrium}}
\vspace{-0.1cm}
\end{figure*}

In the present paper, we adopt the quark masses $m_u=m_d=0$ MeV, $m_s=100$ MeV, and choose two sets of parameters for the bag model, \ie $B^{1/4}=136\ {\rm MeV},\ \alpha_s=0.5$ and $B^{1/4}=140\ {\rm MeV},\ \alpha_s=0$. Both parameter sets result in absolute stability of strange quark matter. 

In Fig.~\ref{fig:bag}, we show the cold $\beta$-equilibrated EoS of quark matter, the mass-radius relation, and the tidal deformability of the corresponding quark stars. Note that here we use the $Y_e=0$ data as approximation of the equilibrium EoS.
The electron fraction $Y_e$ remains extremely low ($\sim 10^{-3}-10^{-6}$ in the bag model) at $\beta$-equilibrium, as the strange quarks take over the role of electrons in maintaining charge neutrality. 
Both models are able to produce NS maximum masses exceeding $2.0 \,\Msun$, as well as reasonable radii and tidal deformabilities that are consistent with current GW and NICER constraints~\citep{2017PhRvL.119p1101A, 2019ApJ...887L..24M, 2019ApJ...887L..21R, 2021ApJ...918L..28M, 2021ApJ...918L..27R}.

\subsection{Deviation from $\beta$-equilibrium}
The FLT-extension method developed so far assumes constant scalar potentials and Dirac effective masses, which are pre-determined as model parameters. We adopt the current quark masses $m_u=m_d=0$ MeV, $m_s=100$ MeV, while setting $M_{\rm diff}$ as a free parameter. However, for negative values of $M_{\rm diff}$, the effective masses of $u$ and $d$ quarks can become negative. Mathematically, negative effective masses are equivalent to their corresponding positive counterparts, as they always appear in the squared form. In practice, we impose a lower bound of zero on 
$M_i$'s; for instance, when $M_{\rm diff}=-20$ MeV, the resulting masses are $M_u=M_d=0$ MeV and $M_s=80$ MeV. 

In Fig.~\ref{fig:out_of_equilibrium}, we compare 
pressures of the FLT extension  
with those of the bag model at $Y_e=0.2$ and $0.5$ (under cold conditions; $T=0.01$ MeV).
These electron fractions deviate from $\beta$-equilibrium for which $Y_e \approx 0$. Triangular markers represent results of the bag model, while the FLT extensions are depicted by solid ($M_{\rm diff}=0$ MeV) and dashed ($M_{\rm diff}=-20$ MeV) lines. The lower panels show deviations in pressure relative to the bag model. 
Larger deviations in the $Y_e=0.5$ case implies higher pressures compared to the $Y_e=0.2$ case. 

Note that the relative pressure errors between the bag model EoSs and our FLT extensions remain below $10^{-2}$. Such small deviations indicate that the FLT extension effectively captures the out-of-equilibrium properties of the bag model. In fact, with constant effective masses and uniform vector potentials, Eqs.~(\ref{eq:nu})--(\ref{eq:muds}) closely resemble those of the bag model EoSs. The conditions of charge neutrality and baryon number conservation depend on the Fermi momenta, 
leaving them identical for both models. 
In addition, since the vector potential is uniform across all quark flavors and canceled out exactly in $d-s$ weak equilibrium, it has no impact on the flavor composition. Moreover, the effective masses of quarks are much smaller than the Fermi momenta, which dominate the $d-s$ weak equilibrium and lead to similar solutions. This also explains why the effects of $M_{\rm diff}$ on out-of-equilibrium pressure are minimal, as shown in Fig.~\ref{fig:out_of_equilibrium}. 

\subsection{Thermal contributions}
\begin{figure}[t]
\vspace{-0.3cm}
{\centering
\includegraphics[width=0.50\textwidth]{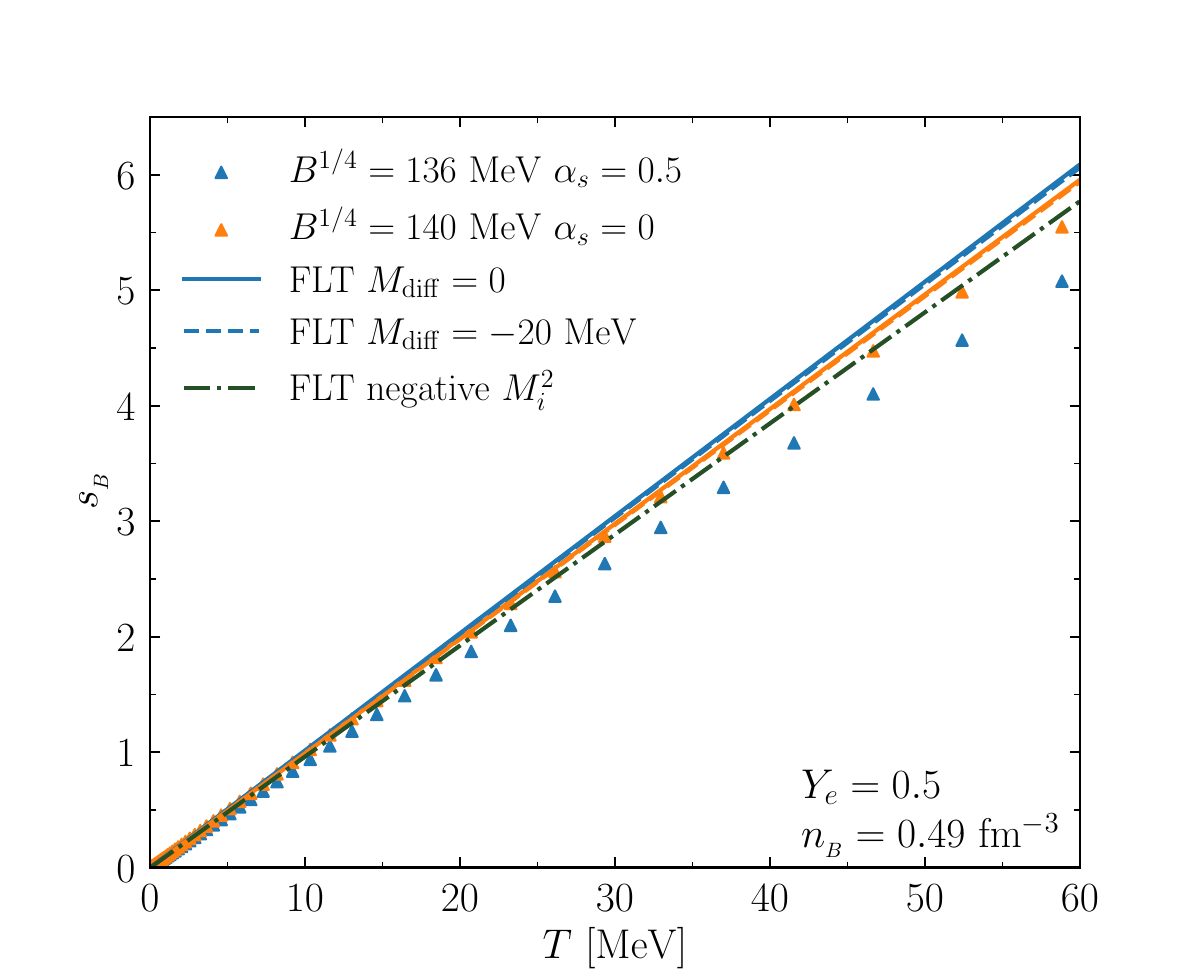}
\caption{ 
Entropy per baryon as a function of temperature at $Y_e=0.5$ and $\nb=0.49$ fm$^{-3}$.
Results from both the bag model EoSs and FLT extensions are shown for comparison, with the same labels as in Fig.~\ref{fig:out_of_equilibrium}. Additionally, we include the result of an FLT model with negative squared Dirac effective masses $M_i^2$; see text for details.}
\label{fig:thermal_T_s}}
\vspace{-0.1cm}
\end{figure}
In the ultradense regions where quark matter resides, quark chemical potentials far exceed  temperatures reached 
in supernovae or BNS mergers. 
As a result, thermal contributions to the EoS can be well approximated by a degenerate gas. In this degenerate limit, entropy exhibits a linear dependence on  $T$, while thermal pressure scales with $T^2$. In FLT, the coefficients of these dependencies are determined by the Landau effective masses, which, in turn, rely on the interactions.

Conventional hydrodynamic formulations evolve rest mass density, momentum, and energy, while entropy is not directly involved in the evolution (see Refs.~\citep{2018PhRvD..97h3014H, 2019MNRAS.490.3588M} for the formulation that evolves entropy). It only plays a role in the primitive variables recovery, where it is independently tracked and used as a backup strategy when the standard recovery fails. Nevertheless, entropy depends to leading order linearly on $T$, with its slope carrying information about effective masses; see \eg Eq.~(\ref{eq:s_bag_alphas}) for the bag model and Eq.~(\ref{eq:entropy}) for the FLT extensions.
In Fig.~\ref{fig:thermal_T_s}, we calculate the entropy per baryon as a function of temperature for both the bag model and FLT-extended EoSs and compare their differences. The entropy per baryon is shown for a slice at $Y_e=0.5$ and $\nb=0.49$ fm$^{-3}$. 
The linear dependence of entropy per baryon on temperature is evident for both models. However, deviations from linearity become noticeable at high enough temperatures for the bag model, particularly for the case with $B^{1/4}=136$ MeV, $\alpha_s=0.5$. In fact, negligible and imperceptible deviations are also present in the FLT extensions, despite the linear dependence being analytically expressed in Eq.~(\ref{eq:entropy}). Due to the temperature dependence of the components, the slope of entropy per baryon can vary with temperature. The non-vanishing coupling constant $\alpha_s$ in the bag model introduces an $O(T^3)$ term in the entropy expression, leading to a larger deviation from linearity (blue triangular markers).

We note that the parameter set with $B^{1/4}=136$ MeV, $\alpha_s=0.5$ leads to smaller entropy per baryon at high temperatures. According to the expression for entropy in Eq.~(\ref{eq:entropy}), the case with the smallest entropy corresponds to the smallest Dirac effective masses. However, it turns out that all the resulting entropies in bag model EoSs are even smaller than the minimum value obtained in the FLT case. 
One reason is that higher-order terms in $T$ contribute negatively. 
Another reason is the effects of the coupling constant $\alpha_s$. To illustrate this, consider a simplified case for the bag model with a non-vanishing $\alpha_s$ and zero quark mass, where the number density can be expressed as
\begin{eqnarray}
  \label{eq:nb_simp}
  n_{_i} = \frac{g \,p_{F_i}^3}{6 \pi^2} - \frac{2\alpha_s}{\pi} \frac{\mu_i^3}{\pi^2} = (1-\frac{12\alpha_s}{g \pi})\frac{g \,p_{F_i}^3}{6 \pi^2}.
\end{eqnarray}
Compared to the FLT expression of number density $n_i=g \,p_{F_i}^3/(6\pi^2)$, presence of $\alpha_s$ in the bag model scales down the Fermi momentum and chemical potential, leading to a smaller Landau effective mass and entropy slope. 
The coupling constant $\alpha_s$ implies an attractive interaction, as it contributes negatively to the pressure. In nucleonic models, such attractive effects are accounted for by the scalar potential $M_{\rm diff}$, which lowers the Dirac effective masses and chemical potentials. However, in quark matter, these effects are difficult to reproduce via the scalar potential alone due to the small current quark masses, which make the reduction of the Dirac effective mass less effective. Since the Fermi momenta dominate the Landau effective masses, influences of the scalar potential are negligible as seen in Fig.~\ref{fig:thermal_T_s}. To better capture this attractive interaction, we introduce negative squared Dirac effective masses $M_i^2$, leading to a modified expression for the chemical potential
\begin{eqnarray}
  \label{eq:neg_em}
  \mu_i = \sqrt{p_{F_i}^2 - M_i^2} + V.
\end{eqnarray}
This approach enhances the applicability of FLT by allowing the Landau effective masses to extend below the previously established lower limit $\min \, (m^\ast_{i}) = p_{F_i}$. Indeed, as shown in Fig.~\ref{fig:thermal_T_s}, the FLT model with negative $M_i^2$ where we set $M_u^2=M_d^2=M_s^2=-100^2$ MeV$^2$, yields a reduced slope and lower entropy. However, it should be noted that negative values of $M_i^2$ do not have a clear physical interpretation; rather, they serve as a mathematical trick to reduce the Landau effective mass. This can be regarded as an effective treatment of the residual attractive interaction.
\begin{figure}[t]
\vspace{-0.3cm}
{\centering
\includegraphics[width=0.50\textwidth]{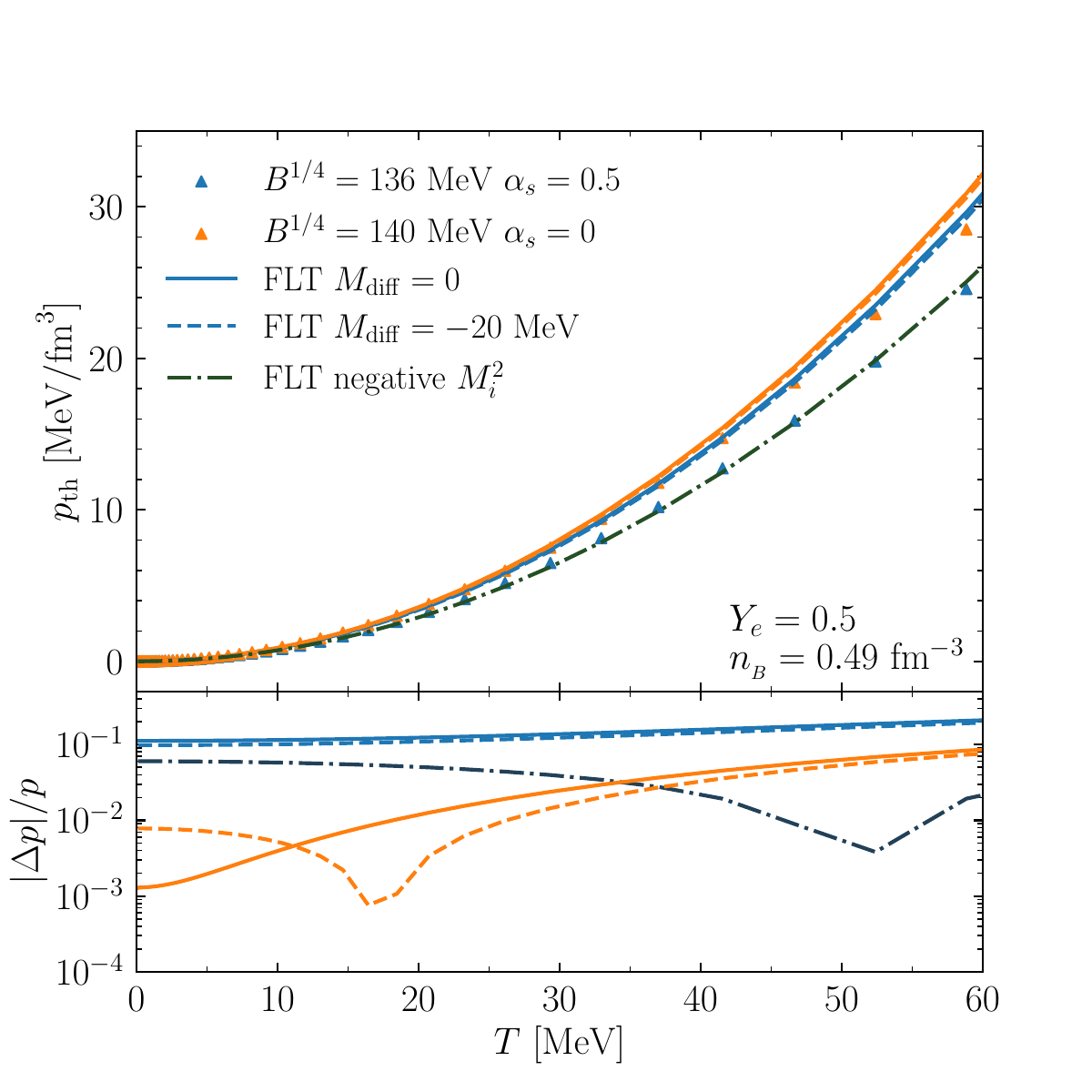}
\caption{Thermal pressures of the bag model and FLT-extended EoSs (upper panel) and their relative errors (lower panel).
Labels are the same as those in previous figures.}
\label{fig:thermal_T_pth}}
\vspace{-0.1cm}
\end{figure}

We also present thermal pressures of the same models in 
Fig.~\ref{fig:thermal_T_pth} with the lower panel showing relative errors between the FLT extensions and the bag model EoSs. 
Similar to the entropy per baryon in Fig.~\ref{fig:thermal_T_s}, FLT-extended EoSs with positive squared Dirac effective masses $M_i^2$ 
give rise to a higher thermal pressure compared to the bag model ones. For the $B^{1/4}=140$ MeV, $\alpha_s=0$ case, the relative errors remain below $10^{-1}$, indicating that FLT extensions can accurately reproduce the thermal pressure. However, when $\alpha_s$ becomes non-zero, the errors increase to around $10^{-1}$. It is remarkable that the effects of $\alpha_s$ can be captured by the FLT model with negative $M_i^2$, leading to a significant reduction in errors to below $10^{-1}$.

\begin{figure}[t]
\vspace{-0.3cm}
{\centering
\includegraphics[width=0.50\textwidth]{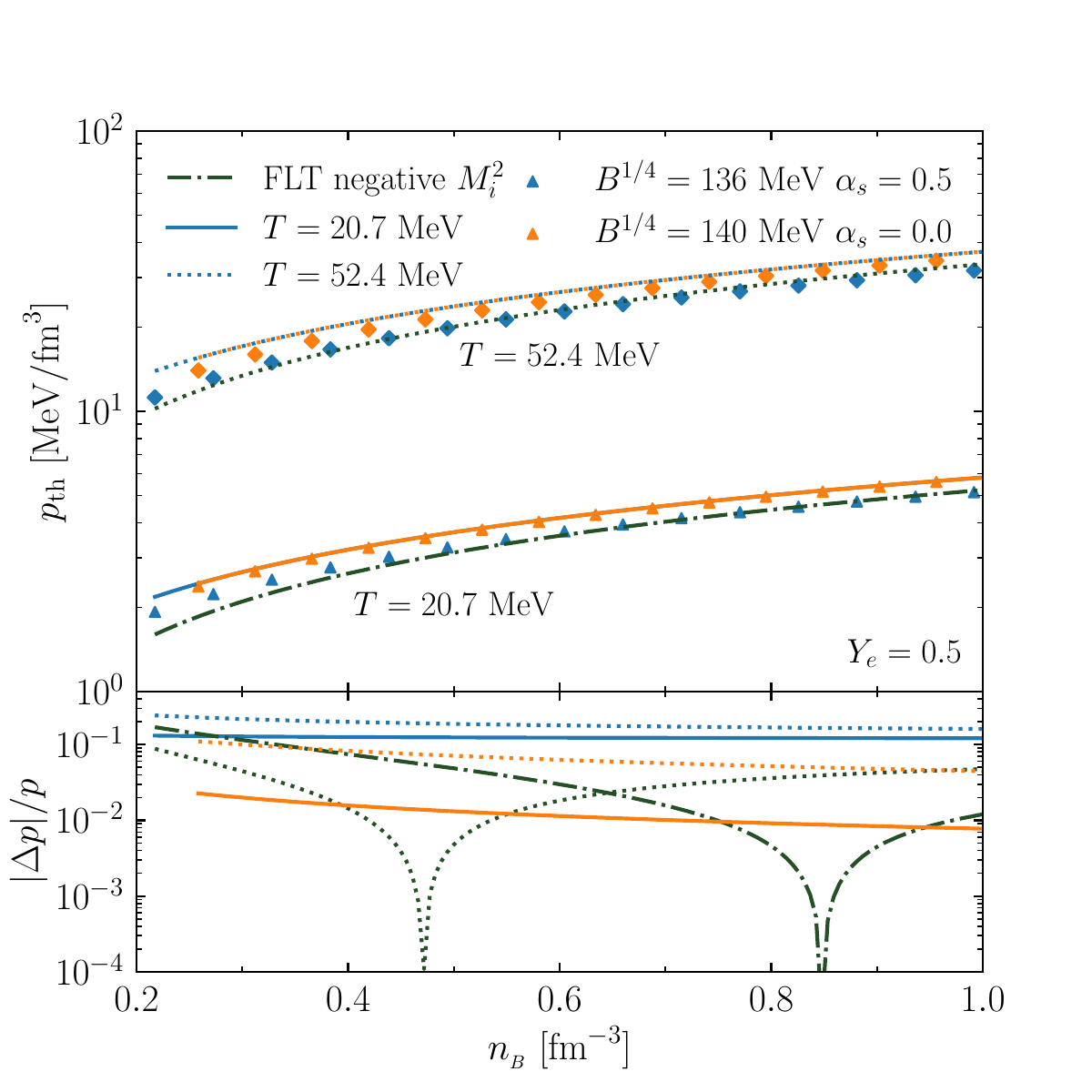}
\caption{Thermal pressure as a function of number density at $Y_e=0.5$. 
Triangular and diamond markers denote the results of the bag model (with the same parameter sets as shown in previous figures), 
for $T=20.7$ MeV and $T=52.4$ MeV, 
respectively, 
whereas solid and dotted lines represent the FLT results.
For all FLT extensions, we set $M_{\rm diff}=0$ MeV, except for the one with negative $M_i^2$ (dash-dotted).}
\label{fig:thermal_nb_pth}}
\vspace{-0.1cm}
\end{figure}

Finally, we show in Fig.~\ref{fig:thermal_nb_pth} the thermal pressure as a function of the number density $\nb$ at $Y_e=0.5$. 
Triangular
and diamond markers correspond to the results of the bag model for $T =
20.7$ MeV and $T = 52.4$ MeV, respectively, while the FLT extended EoSs are represented by solid and dotted lines for these two temperatures. We set $M_{\rm diff}=0$ MeV for all FLT extensions, except for the case with negative squared Dirac effective masses $M_i^2$ (black dash-dotted). 

Notably, differences between the FLT extensions with $B=136$ MeV, $\alpha_s=0.5$ and $B=140$ MeV, $\alpha_s=0$ are imperceptible, leading to almost overlapping lines in this figure. In FLT, effects of the bag constant $B$ are absorbed by the vector potential $V(\nb)$ and do not influence the Landau effective masses, which solely determine the thermal properties. Meanwhile, effects of the coupling constant $\alpha_s$ cannot be captured unless negative $M_i^2$ is introduced. By
lowering Landau effective masses, negative $M_i^2$ reduces the thermal pressure, significantly decreasing the discrepancies with the bag model that includes nonzero $\alpha_s$.

We note that the errors for $B=136$ MeV, $\alpha_s=0.5$ remain nearly constant throughout the density range. As seen in Eq.~(\ref{eq:nb_simp}), the coupling constant $\alpha_s$ effectively introduces a scaling factor on the Fermi momenta and Landau effective masses. This scaling results in a consistent deviation between the FLT and bag model predictions, leading to nearly constant errors across different densities.

In general, our present form of the FLT extension successfully reproduces results of bag models. This agreement benefits from the simple structure of bag models, where the contribution of the ideal Fermi gas dominates, and from the fact that the assumptions and simplifications adopted in Eqs.~(\ref{eq:pe_quarks})–(\ref{eq:q_potentials}) agree well with it. However, 
it is worth noting that these assumptions are not universally suitable for all models of quark matter, particularly in cases where the Dirac effective masses are not trivially constant and the differences between $u$ and $d$ quarks can 
be crucial~\citep{2017EPJC...77..512C, 2019PhRvD.100j3012C, 2021EPJC...81..569C, 2025PhRvD.111j3037P, 2025arXiv250814795Y}.

\section{Treatment of the Phase transition}
\label{sec:PT}
Another important application of the FLT extension is 
to construct 3D hybrid EoS tables with a first-order phase transition (PT) from hadronic to quark matter. 
A first-order PT can induce a sudden softening of the EoS, leading to various observable effects in BNS mergers and CCSNe~\citep{2019PhRvL.122f1101M, 2019PhRvL.122f1102B, 2020PhRvL.124q1103W, 2009PhRvL.102h1101S,2018NatAs...2..980F,2020PhRvL.125e1102Z,2021ApJ...911...74Z,2022ApJ...924...38K,2022PhRvD.106l3037Z}. Typically, a 1D EoS with PT can easily be 
obtained using parameterized approaches (\eg the constant-sound-speed parameterization~\cite{Alford:2013aca}). These 
approaches are flexible to 
cover a vast range in the parameter space, from very soft to very stiff EoSs, and hence have been widely used in parameter estimation and inference from observations~\citep{2020ApJ...904..103M, 2021PhRvD.104f3003L, 2021PhRvC.103d5808D, 2022PhRvC.105c5808D, 2023PhRvD.108d3013E}. However, extending these methods to thermal and out-of-equilibrium scenarios is challenging due to the lack of a microphysical description. This limitation restricts their applicability for systematic investigations of processes in which thermal effects and neutrinos play a significant role~\citep{2023PhRvD.108f3032B}.

In this section, we apply the FLT extension to the constant-sound-speed (CSS) parametrization of quark matter EoS~\cite{Alford:2013aca}, and match them with the DD2 EoS table of hadronic matter~\citep{2010PhRvC..81a5803T, 2010NuPhA.837..210H} to generate 3D EoS tables under Maxwell construction with a self-consistent PT. We use $M_{\rm diff}=0$ MeV in the FLT extension for the remainder of this paper. 

\subsection{Constant-sound-speed parameterization}
In CSS parameterization~\cite{Alford:2013aca}, 
the speed of sound in quark matter is considered density-independent which remains a reasonable approximation for various quark models.
The energy density as a function of pressure can be expressed as
\begin{eqnarray}
  \label{eq:css}
  \ep & = & \ep_0 + \frac{p}{\csq},
\end{eqnarray}
where $\csq$ is the speed of sound squared and a free parameter. The constant $\ep_0$ is determined by the end density of PT (\ie the onset density of the pure quark phase) and its associated pressure. Other variables can be derived by applying the first law of thermodynamics
\begin{eqnarray}
  \label{eq:css_mu}
  \mu_{\rm B} & = & \left[\frac{(1+\kappa)\,p+\ep_0}{A}\right]^{1/(1+\kappa)}, \\
  \label{eq:css_nb}
  \nb & = & A^{1/(1+\kappa)} \left[\ep_0 + (1+\kappa)\,p\right]^{\kappa/(1+\kappa)},
\end{eqnarray}
where we use $\kappa=1/\csq$ for simplicity, and the constant $A$ would also be determined by the chemical potential at the PT.

For any hadronic matter EoS, the CSS parametrization can accommodate a PT at any chosen PT onset density. Given a hadronic EoS such as DD2, we can set the onset number density of the PT $n^h_{\mathrm{PT}}$ to any desired value. The corresponding pressure and chemical potential at this density serve as the PT variables $p_{\mathrm{PT}}$ and $\mu_{\mathrm{PT}}$, while the corresponding energy density is denoted by $\ep_\mathrm{PT}^h$. The constant $A$ is then determined by these two variables as
\begin{eqnarray}
  \label{eq:css_A}
  A & = & \frac{(1+\kappa)\,p+\ep_0}{\mu_\mathrm{PT}^{1+\kappa}}.
\end{eqnarray}
Additionally, the energy density discontinuity
$\Delta \ep$ of the PT, once specified, determines $\ep_0=\ep_\mathrm{PT}^h+\Delta \ep-\kappa\, p_{\mathrm{PT}}$. 
Finally, the quark matter EoS is fully defined after specifying the sound speed. In summary, constructing a 1D hybrid EoS with PT for any given hadronic EoS in CSS parametrization requires only three parameters: $n^h_{\mathrm{PT}}$, $\Delta \ep$ and $\csq$. In the remainder of this paper, we use $\csq=2/3$, $n_{\mathrm{PT}}^{h}=0.2$\,fm$^{-3}$ and $\Delta \ep=150$ MeV/fm$^{-3}$ as an example.

\subsection{Conditions of hadron-quark phase transition}
To meet the stability criteria for a first-order PT from hadronic to quark matter, the system needs to maintain
\begin{eqnarray}
  \label{eq:pt_conditions}
  T^{q} = T^{h},\quad p^{q} = p^{h},\quad \mu^{q} = \mu^{h}.
\end{eqnarray}
Here we use superscripts $q$ and $h$ to denote the variables of the quark phase and the hadronic phase, respectively. These three conditions correspond to the thermal, mechanical, and chemical equilibrium of PT. In the present framework of constructing the 3D EoS table, thermal equilibrium is always satisfied, as we search for locations of the PT at each $T$ and $Y_e$ slice. The pressure and mechanical equilibrium can be straightforwardly applied without ambiguity. However, the chemical equilibrium could be more complex due to involvement of multiple chemical potentials and components in both phases.
\begin{figure}
\vspace{-1.0cm}
  {\centering
  \includegraphics[width=0.49\textwidth]{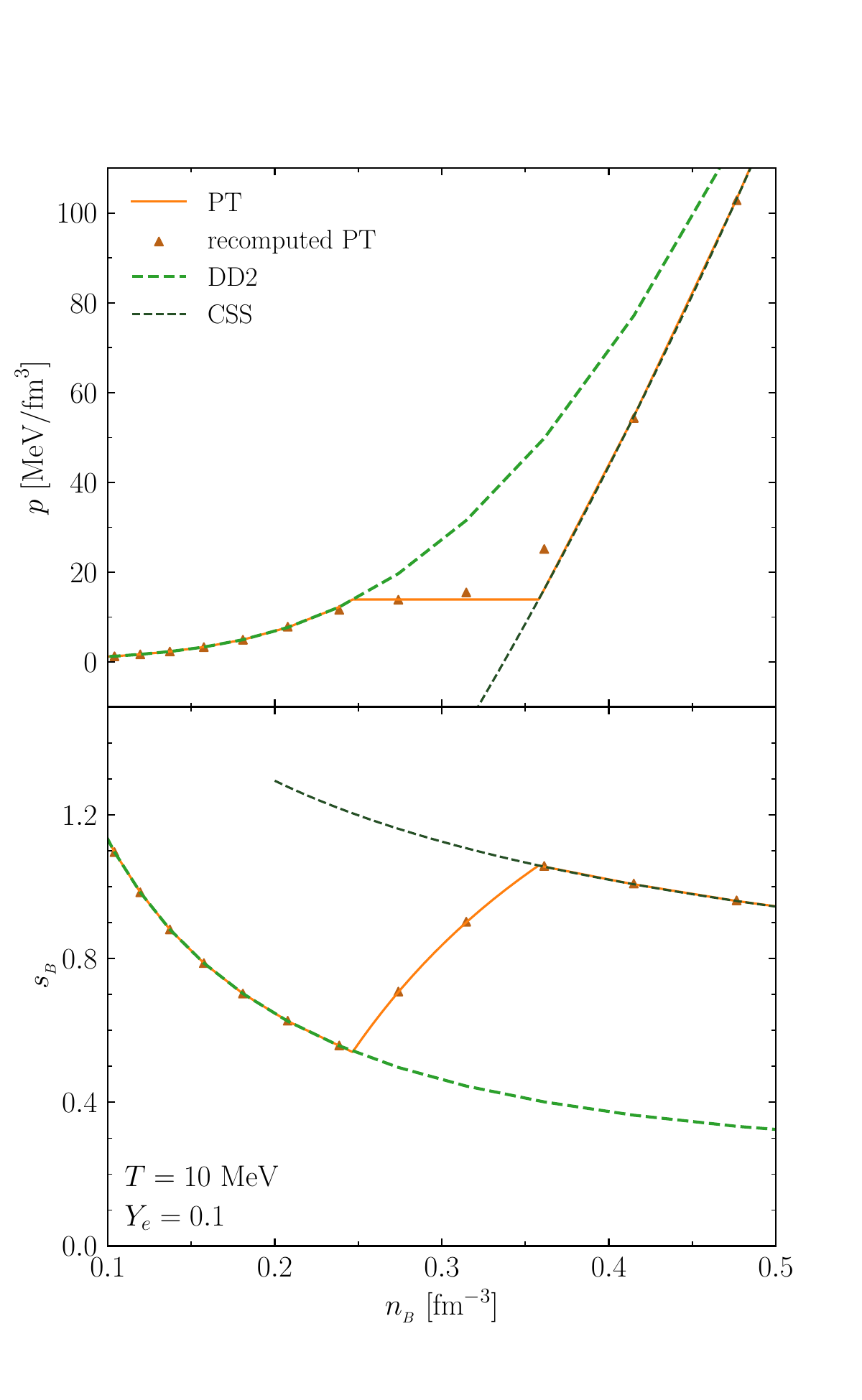}
  \caption{Pressure and entropy per baryon as functions of number density around the PT region.
  The dashed lines represent the purely hadronic (green, DD2) and quark matter (black, CSS) EoS, respectively, while the solid orange line denotes the hybrid EoS with a PT (constant pressure and entropy as described in Eq.~(\ref{eq:entropy_pt}) within the PT region). To ensure thermodynamic consistency, pressure and entropy per baryon must be recomputed by taking the derivatives of Helmholtz free energy. These recomputed quantities are represented by triangular markers.}
\label{fig:pt_p_s}}
\vspace{-0.1cm}
\end{figure}
In Eq.~(\ref{eq:mu_b}), the energy differential in the quark phase is expressed in terms of the number density and chemical potential. 
Similarly for the hadronic phase, we have
\begin{eqnarray}
  \label{eq:mu_b_h}
  d\ep = \mu_n d\nb + (\mu_p+\mu_e-\mu_n) d n_e.
\end{eqnarray}
Note that the second term vanishes at $\beta$-equilibrium for both Eqs.~(\ref{eq:mu_b}) and (\ref{eq:mu_b_h}). Therefore, the chemical equilibrium condition of PT becomes
\begin{eqnarray}
  \label{eq:pt_chem_beta}
  \mu_n = \mu_u + 2\mu_d.
\end{eqnarray}
However, in fixed $Y_e$ cases, the second term of Eq.~(\ref{eq:mu_b_h}) plays a role in PT. In this scenario, the chemical potential of chemical equilibrium is expressed as
\begin{eqnarray}
  \label{eq:pt_chem_condition}
  \mu_{\rm ce} = \left(\frac{\partial \ep}{\partial \nb}\right)_{s_v, Y_e},
\end{eqnarray}
and the condition now reads
\begin{eqnarray}
  \label{eq:pt_chem_ye}
  \mu_n + (\mu_p+\mu_e-\mu_n)Y_e= \mu_u + 2\mu_d + (\mu_u+\mu_e-\mu_d)Y_e. \nonumber \\
\end{eqnarray}
The location of PT for each $T$ and $Y_e$ can be determined by Eq.~(\ref{eq:pt_conditions}). This allows us to identify the onset and end densities of the PT, as well as the chemical potential and pressure at the PT. However, other thermodynamic quantities remain unknown. To fix this we apply the Maxwell relation within the PT region. Specifically,
\begin{eqnarray}
  \label{eq:maxwell}
  -\frac{\partial s}{\partial \rho} \biggr|_T = \frac{1}{\rho^2} \frac{\partial p}{\partial T} \biggr|_\rho,
\end{eqnarray}
where $\rho=\nb \mb$ refers to the rest mass density and $s=s_v/\rho$ denotes the specific entropy. Within the PT region, pressure is independent of density, therefore $(\partial p/\partial T) |_\rho$ remains constant for a given $T$. The specific entropy can then be expressed as
\begin{eqnarray}
  \label{eq:entropy_pt}
  s = \frac{1}{\rho} \frac{\partial p}{\partial T} \biggr|_\rho + C,
\end{eqnarray}
where $C$ is an integral constant and will be determined by the hadronic or quark entropy at the onset or end density of the PT.

In practice, $(\partial p/\partial T) |_\rho$ at the PT can be evaluated through numerical derivative of the PT pressure. The constant $C$ is then determined by applying one of the PT boundary conditions (either at the onset or the end). However, due to numerical errors in locating the PT, the entropy may become discontinuous at the other boundary which is unphysical.
To resolve this issue, we apply both boundary conditions to determine these two constants (both $C$ and $\partial p/\partial T$), ensuring continuity of the entropy.

To illustrate, we calculate the pressure and entropy per baryon, $s_{\rm B}=s\,\mb$, around the PT region for $T=10$ MeV, $Y_e=0.1$, 
and the results are shown in Fig.~\ref{fig:pt_p_s}. 
The hadronic DD2 EoS and quark CSS EoS are represented by dashed lines, while the hybrid EoS with PT is shown as the orange solid line. Note that the recomputed entropy or entropy per baryon in the PT region follows closely the analytical relation Eq.~(\ref{eq:entropy_pt}). 
\begin{figure}[t]
\vspace{-0.3cm}
{\centering
\includegraphics[width=0.50\textwidth]{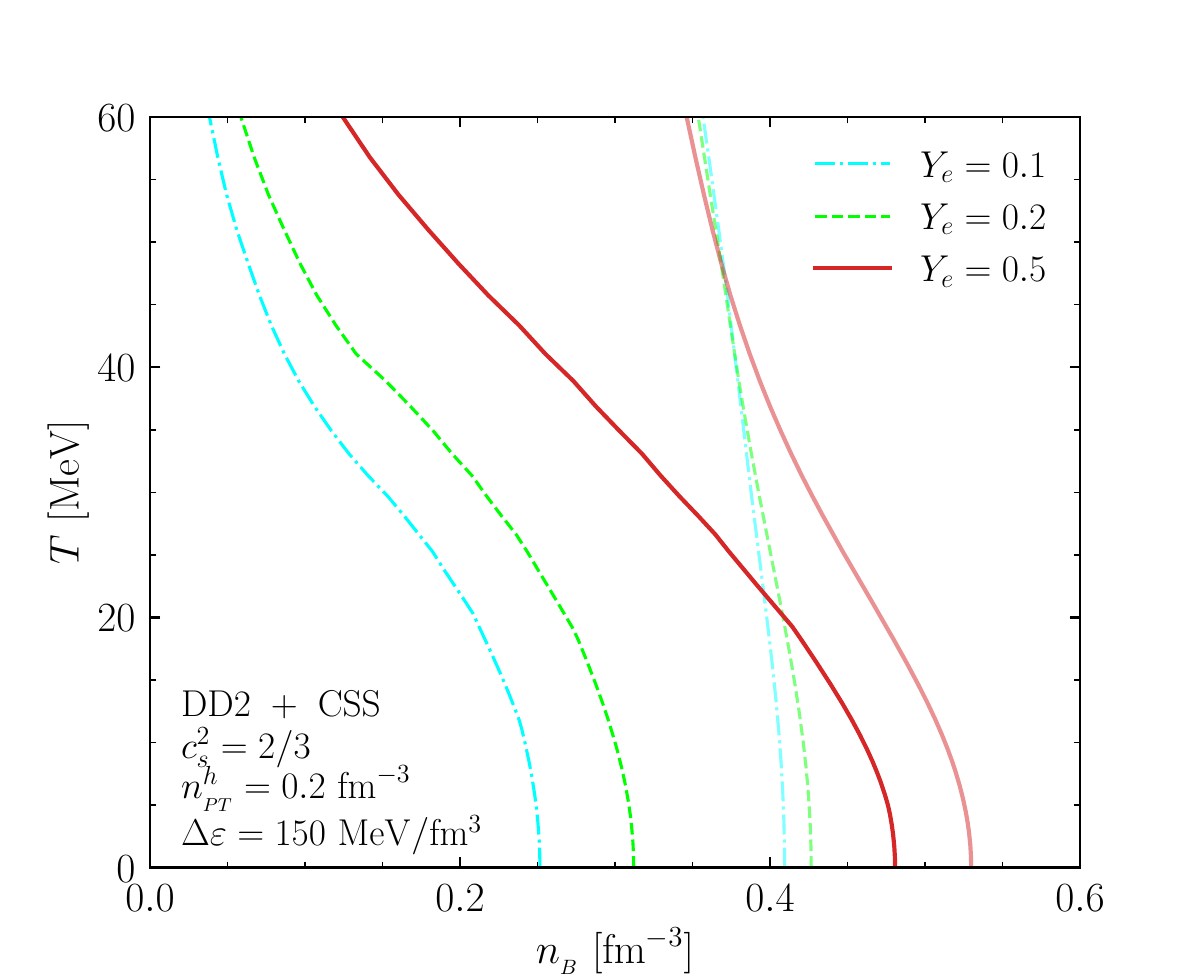}
\caption{
Onset and end densities (horizontal axis) of the PT with varying temperatures (vertical axis) at fixed $Y_e=0.1$, $Y_e=0.2$ and $Y_e=0.5$, represented by dash-dotted, dashed, and solid lines, respectively. The hadronic matter is modeled by the DD2 EoS, while the quark matter is described using the CSS parameterization with $\csq=2/3$, $n_{\mathrm{PT}}^h=0.2\ {\rm fm}^{-3}$ and $\Delta \ep=150$ MeV/fm$^{-3}$.}
\label{fig:pt_positions}}
\vspace{-0.1cm}
\end{figure}

\subsection{3D hybrid equation of state with phase transition}
By identifying the intersection of the two $\mu_{\rm ce}$-$p$ functions for hadronic and quark matter, we can determine the phase boundaries. It should be noted that the CSS parameters $n^h_{\mathrm{PT}}$ and $\Delta \ep$ characterize the PT only at zero temperature and under weak equilibrium conditions, \ie 1D hybrid EoS. 
In contrast, boundaries of the PT in the full 3D EoS table are determined self-consistently within our EoS extension framework. In principle, the 3D hybrid EoS and PT parameters are generated self-consistently, ensuring that all thermodynamic relations are satisfied. 
However, in practice, numerical errors can introduce deviations that potentially cause problems in hydrodynamic simulations, which are highly sensitive to these relations. 
These problems become even more severe in the case of PTs where discontinuities are present. Therefore, additional treatments are necessary to maintain consistency for all slices of $Y_e$.

Starting from the first law of thermodynamics, we have
\begin{eqnarray}
  \label{eq:firstlaw}
  d\epsilon = Tds + \frac{p}{\rho^2} d\rho,
\end{eqnarray}
where $\epsilon$ represents the specific internal energy. This differential form implies the relations between partial derivatives
\begin{eqnarray}
  \label{eq:thermo_relations1}
  p & = & \rho^2 \frac{\partial \epsilon}{\partial \rho}\biggr|_T + T \frac{\partial p}{\partial T}\biggr|_\rho , \\
  \label{eq:thermo_relations2}
  \frac{\partial \epsilon}{\partial T}\biggr|_\rho & = & T \frac{\partial s}{\partial T}\biggr|_\rho, \\
  \label{eq:thermo_relations3}
  \frac{\partial s}{\partial \rho}\biggr|_T & = & -\frac{1}{\rho^2} \frac{\partial p}{\partial T}\biggr|_\rho.
\end{eqnarray}
All of these relations must be strictly satisfied by the generated EoS table. To ensure this, we use the Helmholtz free energy~\citep{2021ApJ...913...72J},
\begin{eqnarray}
  \label{eq:free_energy}
  F = \epsilon - Ts,
\end{eqnarray}
and recompute the pressure and entropy by taking the derivatives of $F$ with respect to $\rho$ and $T$
\begin{eqnarray}
  \label{eq:recompute_ps}
  p = \rho^2 \frac{\partial F}{\partial \rho}\biggr|_T, \quad 
  s = -\frac{\partial F}{\partial T}\biggr|_\rho.
\end{eqnarray}
One can easily verify that the recomputed pressure and entropy satisfy all the relations~(\ref{eq:thermo_relations1})--(\ref{eq:thermo_relations3}).
\begin{figure}[t]
\vspace{-0.3cm}
{\centering
\includegraphics[width=0.50\textwidth]{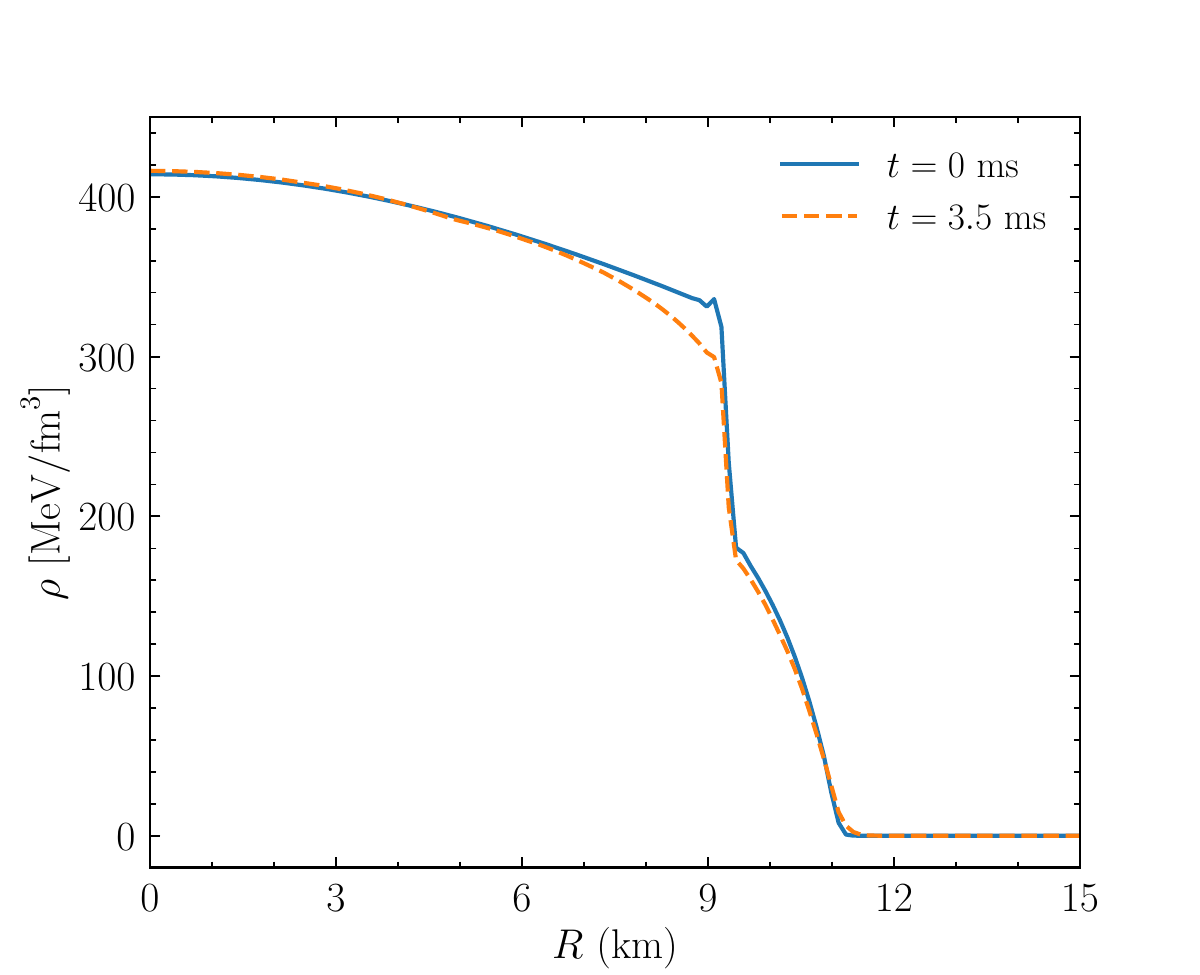}
\caption{
Density profiles of the TOV star at the initial time (blue) and after $3.5$ ms evolution (orange). Note the differences near the PT region, which arise from the use of the 1D high-resolution EoS in the initial data and the sliced 3D EoS during evolution. See text for further discussion.}
\label{fig:tov_1d}}
\vspace{-0.1cm}
\end{figure}
In Fig.~\ref{fig:pt_p_s}, we present an example of constructing hybrid EoSs with a PT, showing both the original (solid, orange lines) and recomputed (triangular markers) pressure and entropy per baryon as functions of the number density $\nb$. 
Interestingly, the recomputed pressure deviates slightly from the plateau of the PT and is smoothed out by taking derivatives of the free energy with respect to $\rho$. This recomputation procedure ensures thermodynamic consistency by eliminating discontinuities in the pressure.

Finally, in Fig.~\ref{fig:pt_positions} we present the onset and end densities of the PT at fixed $Y_e=0.1$, $Y_e=0.2$ and $Y_e=0.5$ with varying temperatures. 
The hadronic matter is modeled by the DD2 EoS, while the quark matter is described by the CSS parameterization with $\csq = 2/3$, $n^h_{\mathrm{PT}}=0.2\ {\rm fm}^{-3}$ and $\Delta \ep=150$ MeV/fm$^{-3}$. Note that $n^h_{\mathrm{PT}}$ in the CSS parameters refers to the onset density of PT at $\beta$-equilibrium. Out-of-equilibrium effects shift the chemical potential $\mu_{\rm ce}$ to higher values, leading to a 
higher onset density than $n^h_{\mathrm{PT}}$ at zero temperature for all $Y_e$ cases. In general, a larger deviation from the equilibrated $Y_e$ results in a delayed onset. These effects are similar to those reported in Ref.~\citep{2009PhRvL.102h1101S}, where a Gibbs construction was applied for the PT.
On the other hand, finite temperatures shift the onset density towards lower values, which plays an important role in the post-merger evolution of BNS mergers; 
see detailed discussion in Ref.~\citep{2023PhRvD.108f3032B}.

\section{General relativistic hydrodynamic simulations}
\label{sec:simulation}
The ultimate goal of constructing the full 3D EoSs utilizing the FLT extension method is its application in general relativistic hydrodynamic (GRHD) simulations, especially in scenarios involving first-order PTs that introduce density discontinuities within neutron stars. Such discontinuities require a thermodynamically consistent EoS table to ensure the stability and robustness of hydrodynamic evolutions. 
To demonstrate the effectiveness of our approach, in particular the robustness of treating PT in our 3D EoS construction, in this section, we present two representative examples of GRHD simulations detailed as follows.  

\subsection{Evolving TOV star with a phase transition}
The TOV-star evolution is the simplest scenario for testing GRHD simulations of NSs (in particular for hybrid stars with a first-order PT), where we impose the static TOV solution as the initial condition and evolve the spacetime and hydrodynamics. Specifically, we perform evolution of a $1.4 \,\Msun$ hybrid star with PT triggered at around 1.5 times nuclear saturation density using the \texttt{Frankfurt-IllinoisGRMHD (FIL)} code~\citep{2019MNRAS.490.3588M} for $\sim 3.5$ ms, until the star reaches a stable configuration. 
Density profiles of the star are shown in Fig.~\ref{fig:tov_1d}, where we compare the initial configuration with the state after $3.5$\,ms of evolution.

The initial condition is obtained by solving the static TOV equations using a high-resolution 1D (cold and $\beta$-equilibrated) EoS table, which allows the PT region to be well resolved. However, for the dynamical GRHD evolution that follows, the full 3D EoS table must be applied for which the PT region is sampled by only a few grid points on the density due to the file size constraint (see Fig.~\ref{fig:pt_p_s}). Furthermore, the recomputation process to a small extent smooths out the density discontinuity, as discussed in the previous section, resulting in a relatively narrower transition region. These effects contribute to the difference observed in density profiles between the initial data (blue solid line) and the evolved solution at $3.5$\,ms (orange dashed line).

Due to the mismatch between the stable configuration of the simulation and the initial data, an oscillation is generated shortly after the beginning of the evolution. This oscillation quickly dissipates, and the system settles back into a new stable state. Importantly, the density discontinuity remains stable and well resolved throughout the simulation, demonstrating the robustness and reliability of the 3D EoS table.

\subsection{Core-collapse supernova simulations}
CCSNe mark the death of massive stars with an initial mass greater than $\sim8-10\,\Msun$, giving birth to NSs (or hybrid and quark stars) and stellar-mass black holes, and they are among the most violent explosive events in the universe. The CCSN explosion mechanism involves all 4 fundamental interactions and has been a mystery in modern astronomy (see, \eg \cite{1990RvMP...62..801B,2012ARNPS..62..407J,2021Natur.589...29B}). A first-order hadron-quark PT offers a possible solution for successfully driving a CCSN explosion that operates in spherically symmetric simulations \cite{2009PhRvL.102h1101S,2018NatAs...2..980F}. As the core of a massive star collapses and forms a protoneutron star (PNS), a stalled accretion shock forms at $\sim100-200$\,km and the PNS continuously gains mass and contracts to become increasingly compact. A successful explosion is launched only if the stalled shock is revived into runaway expansion. During the accretion phase, a first-order PT may take place and lead to the collapse of the PNS. This collapse can be followed by a bounce shock, which can result in the explosion of the outer mantle if sufficiently powerful. The success and failure of the PT-driven explosion depend both on the EoS and progenitor properties \cite{2021ApJ...911...74Z,2022ApJ...924...38K,2022MNRAS.516.2554J}. 

Here, we apply the example EoS to CCSN simulations with several progenitor models to demonstrate its application. We use the 1D GRHD code \texttt{GR1D} augmented with a two-moment neutrino transport scheme \cite{2015ApJS..219...24O}. Neutrinos are evolved with 3 species ($\nu_e$, $\bar{\nu}_e$, $\nu_x$=\{$\nu_\mu$, $\bar{\nu}_\mu$, $\nu_\tau$, $\bar{\nu}_\tau$\}) and 18 energy groups up to 310\,MeV for each species. Interaction rates between neutrinos and matter are solved with the \texttt{NuLib} library, while we utilize effective nucleon chemical potentials from quark chemical potential (cf. Eq.~\ref{eq:pt_chem_beta}) in the quark matter phase. The spatial resolution is 300\,m in the central 20\,km and becomes logarithmically increasing outwards, up to $2\times10^4$\,km with 1000 zones. We use solar-metallicity progenitor models from Ref.~\cite{2018ApJ...860...93S} with initial masses of $16\,\Msun$, $19\,\Msun$ and $24\,\Msun$.
\begin{figure}[t]
\vspace{-0.3cm}
{\centering
\includegraphics[width=0.48\textwidth]{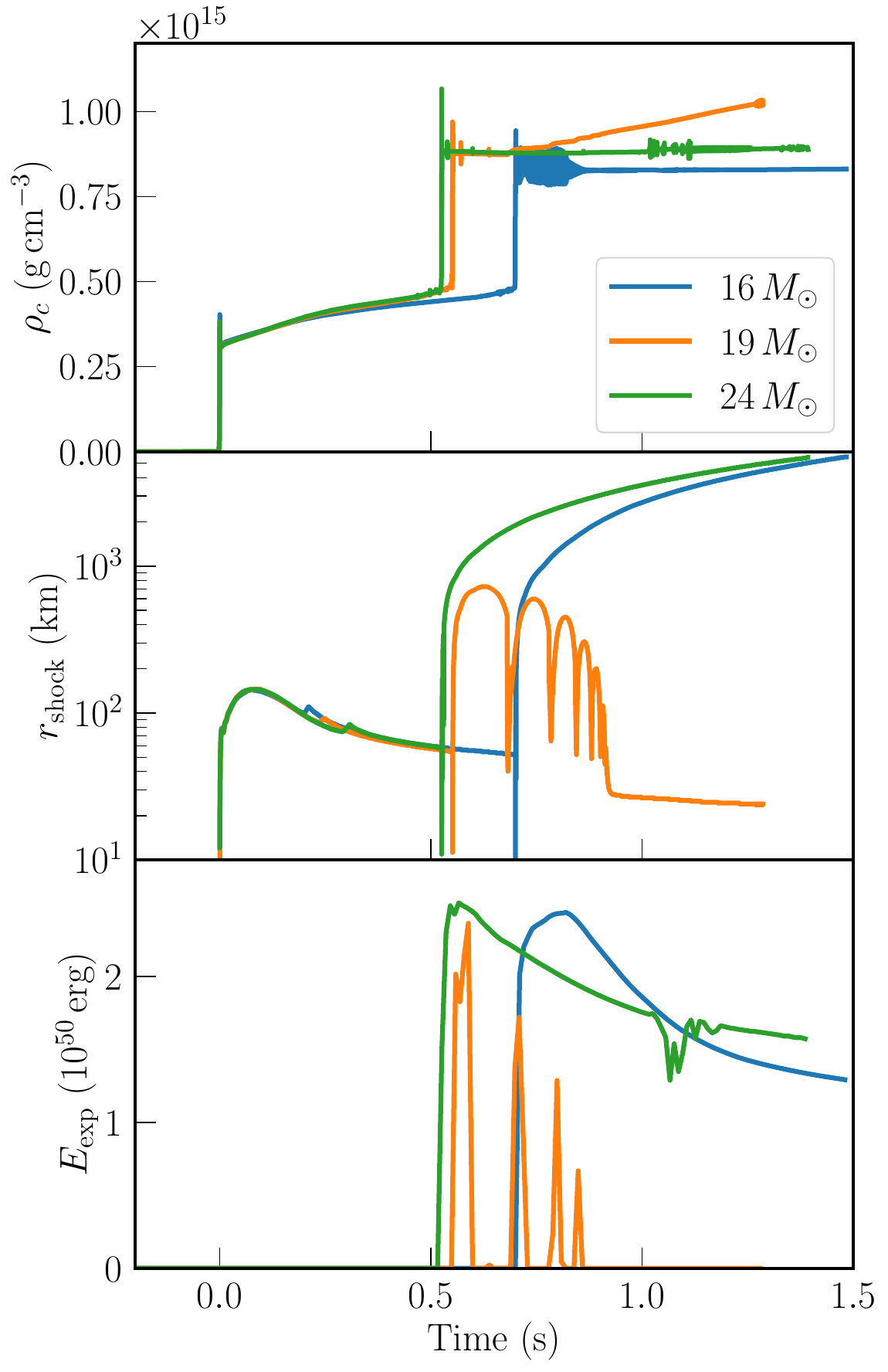}
\caption{Time evolution of the central density (upper panel), shock radius (middle panel) and explosion energy (lower panel) in the 3 CCSN simulations with the example 
hybrid hadron-quark EoS. Note that time zero is defined as the moment of the first core bounce, which occurs due to the stiffening of the nuclear matter equation of state.}
\label{fig:ccsn}}
\vspace{-0.1cm}
\end{figure}

Figure~\ref{fig:ccsn} shows the main quantitative results from the CCSN simulations. About $0.5-0.7$\,s after the first core bounce due to the stiffening of the nuclear EoS (defined as time zero in the figure), the first-order PT occurs at a central density of $\sim5\times10^{14}\,{\rm g\,cm^{-3}}$ and it leads to dynamical collapse of the PNS with a surge of the central density, reaching about $10^{15}\,{\rm g\,cm^{-3}}$. The core bounces due to the stiffening of the quark matter EoS (with respect to the sudden softening at the phase-transition threshold) indicated by the drop of the central density and the expansion of the shock radius. Afterwards, runaway shock expansion occurs in the 16\,$\Msun$ and 24\,$\Msun$ models, while the 19\,$\Msun$ experiences several cycles of shock expansion and fallback. The explosion energy in the 16\,$\Msun$ and 24\,$\Msun$ models remains positive until the end of the simulations when the shock radius has exceeded $\sim5000$\,km, indicating a successful CCSN explosion. The explosion energy reaches a maximum value of $\sim2.5\times10^{50}$\,erg and slowly decreases as the shock propagates through the overburden. We note that the core oscillates for $\sim100$\,ms following its bounce (most prominent in the 19\,$\Msun$ model), with the same period as the TOV-star evolution. This can lead to observable oscillatory features in the neutrino signal, warranting further exploration.
These results qualitatively agree with previous studies employing other dedicated EoSs \cite{2009PhRvL.102h1101S,2018NatAs...2..980F,2021ApJ...911...74Z,2022MNRAS.516.2554J} and thus demonstrate that our 3D EoS extension is pragmatic for realistic astrophysical simulations involving finite-temperature and out-of-equilibrium conditions.
We will conduct a more systematic survey of the PT parameter and progenitor dependence using our 3D EoS extension method in future work. \\

\section{Summary and Conclusions}
\label{sec:cons}

The detection of GWs from BNS mergers along with the simultaneous mass-radius measurements from the NICER mission has significantly advanced our efforts to constrain the cold, dense matter EoS of NSs. However, the cold EoS alone is insufficient to fully describe the interactions among particles in dense matter or comprehensively capture the dynamics of BNS mergers and CCSNe. 
Thermal and out-of-equilibrium effects play important roles in these dynamical evolutions of NSs, where shock and neutrino emissions are typically present. In numerical simulations, a common approach is to approximate the thermal pressure as that of an ideal gas and to neglect the out-of-equilibrium effects and neutrino contributions. Such simplification fails to capture essential ingredients as seen in a full 3D EoS based on physical models, particularly in the presence of a phase transition for which the onset and end densities of the PT depend sensitively on temperature and electron fraction.

In the present work, we developed a method to systematically extend a cold and $\beta$-equilibrated EoS of quark matter to a full three-dimensional version within the framework of Fermi-liquid theory, incorporating both thermal and out-of-equilibrium effects in a self-consistent way. We compared the FLT-extended EoS with the bag model without strong coupling corrections, and found that our FLT extension successfully reproduces contributions from both the temperature $T$ and out-of-equilibrium $Y_e$. Furthermore, the first-order correction from strong interactions in the bag model introduces an attractive effect, yielding a negative contribution to pressure. This additional attraction can be 
effectively 
reproduced in the FLT framework by introducing a negative squared Dirac effective mass by hand, which reduces the Fermi energy and consequently lowers both the pressure and entropy in practice.

We also constructed a 3D hybrid EoS with PT by matching the hadronic DD2 EoS to our FLT-extended quark matter EoS, and tested it through applications to the TOV-star and CCSN simulations. For TOV-star simulation, we evolved the hybrid star for $\sim 3.5$ ms using the initial data computed with a high-resolution 1D EoS with PT. Although our 3D EoS deviates slightly from the 1D version close to the PT point and introduces an initial oscillation, the star quickly settles into the new equilibrium consistent with the 3D EoS. Meanwhile, the density discontinuity of the PT is stably maintained throughout the evolution. We also simulated CCSN explosions driven by the PT for three different progenitor models with initial masses of $16\,\Msun$, $19\,\Msun$ and $24\,\Msun$. 
The 16\,$\Msun$ and 24\,$\Msun$ cases showed positive explosion energy by the end of the simulations, indicating successful explosions. The results of our CCSN simulations are qualitatively consistent with previous studies using alternative EoSs. In summary, both
examples of 
GRHD simulations 
demonstrate the effectiveness and robustness of our 3D EoS construction. 

We 
want to point out that the present form of FLT extension is limited by simplified assumptions of a particle-independent vector potential and also constant Dirac effective masses, both of which are robust in bag models. 
Different 
quark or hadronic matter models 
can, however, involve a more complex structure of SPE in general, for which the density-dependent Dirac effective masses and the isovector potential play essential roles. Moreover, hadronic matter can extend to the low-density and high-temperature region, where the condition for the degeneracy limit $T < \mub-\mb$ breaks down and the FLT becomes invalid. 
In such cases, a new form of temperature dependence in the intermediate temperature region must be introduced to smoothly connect both the low-temperature FLT limit and the high-temperature ideal-gas limit, which could be further investigated in future work. 

After establishing the FLT framework for extending quark matter EoS, 
systematic and comprehensive studies on how the phase transition parameters and the baseline cold EoS properties (such as the strange quark mass and bag constant) can impact simulation results of CCSNe and BNS mergers become feasible. 
More importantly, this generic framework 
holds promise to facilitate 
self-consistent implementation of 
thermal contributions and  neutrino evolutions into future simulations, thereby allowing for reliable predictions relevant to the remnant lifetime, associated gravitational wave spectrum and frequency, properties of the ejecta and concomitant electromagetic emissions, et cetera. We plan to explore these effects in future work. 
\\


\begin{acknowledgments}
The authors thank Jürgen Schaffner-Bielich and Lie-Wen Chen for inspiring discussions, and Constantinos Constantinou for constructive feedback on an earlier version of this manuscript. We gratefully acknowledge Luciano Rezzolla for providing the \texttt{FIL} code.
This work was initiated during the TDLI Workshop ``Dense Matter Equation of State and Frontiers in Neutron Star Physics'' held at the T.D. Lee Institute, Shanghai Jiao Tong University. 
Z.Z. acknowledges support from the National Natural Science Foundation of China (NSFC, No. 12203033) and the China Postdoctoral Science Foundation (No. BX20220207 and 2022M712086).
S.Z. is supported by the National Natural Science Foundation of China (NSFC, No. 12473031, 12393811), the Yunnan Revitalization Talent Support Program--Young Talent project, and by Yunnan Fundamental Research Projects (grant NO. 202501AS070078). 
The work of S.H. was supported by Startup Funds from the T.D. Lee Institute and Shanghai Jiao Tong University. 

\end{acknowledgments}

\section*{DATA AVAILABILITY}
The data that support the findings of this article are
openly available~\citep{data_flt}.


\appendix

\onecolumngrid

\section{The Landau parameters and effective masses}
\label{app:A}

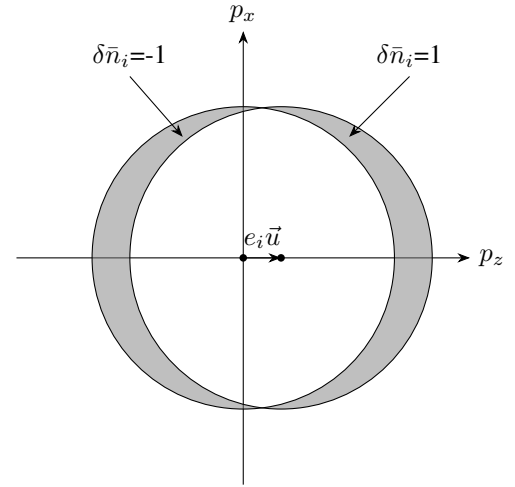
\begin{wrapfigure}{r}{0.4\textwidth}
  \begin{tikzpicture}[> = Stealth]
    \draw[->] (-3,0) -- (3,0) node[right] {$p_z$}; 
    \draw[->] (0,-3) -- (0,3) node[above] {$p_x$}; 

    \fill[gray, fill opacity=0.5, even odd rule] (0,0) circle (2cm) (0.5,0) circle (2cm); 

    \draw (0,0) circle (2cm); 
    \draw (0.5,0) circle (2cm); 
    \fill (0,0) circle (0.05cm); 
    \fill (0.5,0) circle (0.05cm); 
    \draw[->] (0,0) -- (0.5,0) node[midway,above] {$e_i\vec{u}$}; 
    \draw[<-] (-0.8,1.6) -- (-1.5,2.4) node[at end,above] {$\delta \bar{n}_{i}$=-1}; 
    \draw[<-] (1.4,1.6) -- (2.2,2.4) node[at end,above] {$\delta \bar{n}_{i}$=1}; 
  \end{tikzpicture}
  \caption{The distributions of a zero-temperature system before and after a Lorentz boost.}
  \label{fig:boost}
  \vspace{-2.2cm}
\end{wrapfigure}

We recall the definition of Landau effective mass
\begin{eqnarray}
  \label{eq:def_mstar}
  m_i^\ast & = & \left(\frac{d e_i^{(0)}(p_i)}{dp_i}\biggr|_{p_i=p_{F_i}} \right)^{-1} p_{F_i},
\end{eqnarray}
and its connection to the interaction function $f_{i,i'}(\vec{p}_{F_i}, \vec{p}_{F_{i'}})$ can be derived by applying a Lorentz boost to the system. Consider an infinitesimal Lorentz boost with velocity $\vec{u}$ applied to a zero-temperature system (shown in Fig.~\ref{fig:boost}), the energy and momentum of the particles vary as follows:
\begin{eqnarray}
  \label{eq:inf_boost}
  \tilde{e}_i(\vec{\tilde{p}}_i) & = & \gamma_i(e_i + \vec{p}_i\cdot\vec{u}) \approx e_i (\vec{p}_i) + \vec{p}_i\cdot \vec{u} , \\
  \vec{\tilde{p}}_i & = & \vec{p}_i + (\gamma_i-1)(\vec{p}_i\cdot\vec{u})\vec{u}/u^2 + \gamma_i e_i \vec{u}
                 \approx \vec{p}_i + e_i \vec{u}.
\end{eqnarray}
Here, we only keep the first-order terms of $\vec{u}$ and assume that the interactions are Lorentz invariant.

On the other hand, one can consider the boost as an excitation of the system. The distribution functions change from $\bar{n}_{i}=\Theta(p_{F_i}-p_i)$ to $\bar{n}_{i}=\Theta(p_{F_i}-|\vec{p}_i+e_i^{(0)}\vec{u}|)$ after the boost. We can express the variation of the distribution functions as
\begin{eqnarray}
  \label{eq:dni_lb}
  \delta \bar{n}_i =
  \begin{cases}
    1, & {\rm when}\ p_{F_i} < p_i < \sqrt{p_{F_i}^2 + 2\mu_i\vec{p}_i \cdot \vec{u} - \mu_i^2 u^2} ; \\
    -1, & {\rm when}\ \sqrt{p_{F_i}^2 + 2\mu_i\vec{p}_i \cdot \vec{u} - \mu_i^2 u^2} < p_i < p_{F_i} ; \\
    0, & {\rm elsewhere},
  \end{cases}
\end{eqnarray}
where $e_i^{(0)}$ is replaced by $\mu_i$, since the excitation occurs near the Fermi surface, and $e_i^{(0)}$ can be approximated by the Fermi energy $e_{F_i} = \mu_i$. The conditions can then be simplified by neglecting the terms with higher order than $\mathcal{O}(\vec{u})$,
\begin{eqnarray}
  \label{eq:dni_lb2}
  \delta \bar{n}_i =
  \begin{cases}
    1, & {\rm when}\ p_{F_i} < p_i < p_{F_i} + \mu_i u \cos\theta' ; \\
    -1, & {\rm when}\ p_{F_i} + \mu_i u \cos\theta' < p_i < p_{F_i} ; \\
    0, & {\rm elsewhere},
  \end{cases}
\end{eqnarray}
where $\theta'$ denotes the angle between $\vec{p}_i$ and $\vec{u}$.
Note that excitations occur near the Fermi surface (\ie $\delta \bar{n}_i$ is nonvanishing only when $p_i$ is near $p_{F_i}$), and the condition for $\delta \bar{n}_i=1$ ($\delta \bar{n}_i=-1$) simplifies to
$\cos\theta'>0$ ($\cos\theta'<0$). The corresponding width in momentum space (see Fig.~\ref{fig:boost}) is therefore given by $|\mu_i u \cos\theta'|$.

The single particle energy $\tilde{e}_i$ can be obtained by varying the energy functional with respect to the distribution function $\bar{n}_i$
\begin{eqnarray}
  \label{eq:qp_energy}
  \tilde{e}_i = \delta E(\bar{n}_i) / \delta \bar{n}_i = e_i^{(0)} + \sum_{\vec{p}_{i'},i'} f_{i,i'}(\vec{p}_i, \vec{p}_{i'}) \delta \bar{n}_{i'}.
\end{eqnarray}
Consider an excited quasi-particle near the Fermi surface, its energy is identical to the corresponding energy after the boost (both are denoted by $\tilde{e}_i$), as both scenarios describe the same physical phenomenon. Therefore, we have
\begin{eqnarray}
  \label{eq:lb_fl}
  \tilde{e}_i(\vec{\tilde{p}}_i) & = & \tilde{e}_i(\vec{p}_i + e_i \vec{u}) \nonumber \\
  & = & \tilde{e}_i(\vec{p}_i) + \frac{d\tilde{e}_i(\vec{p_i})}{d\vec{p_i}} \cdot e_i \vec{u} \nonumber \\
  & = & e_i^{(0)}(\vec{p}_i) + \sum_{\vec{p}_{i'},i'} f_{i,i'}(\vec{p}_i, \vec{p}_{i'}) \delta \bar{n}_{i'} + \frac{d\tilde{e}_i(\vec{p_i})}{d\vec{p_i}} \cdot e_i \vec{u} \nonumber \\
  & = & \vec{p_i} \cdot \vec{u} + e_i.
\end{eqnarray}
We further set $p_i=p_{F_i}$ and $e_i=e_{F_i}=\mu_i$ in Eq.~(\ref{eq:lb_fl}) and obtain
\begin{eqnarray}
  \label{eq:lb_fl2}
   p_{F_i} u \cos\omega & = & \sum_{\vec{p}_{i'},i'} f_{i,i'}(p_{F_i}, \vec{p}_{i'}) \delta \bar{n}_{i'} + \frac{de_i}{dp_i} \biggr|_{p_{F_i}} \mu_i u \cos\omega,
\end{eqnarray}
where $\omega$ denotes the angle between $\vec{p}_{F_i}$ and $\vec{u}$.

The term $\sum f_{i,i'}(p_{F_i}, \vec{p}_{i'}) \delta \bar{n}_{i'}$ could be evaluated by introducing the expression of $\delta \bar{n}_i$ of Eq.~(\ref{eq:dni_lb}). Without loss of generality, we take the direction of $\vec{u}$ to be along the positive $z$ axis, and set $\vec{p}$ to lie in the xz-plane. The unit vectors of $\vec{p}/|\vec{p}|$ and $\vec{p}'/|\vec{p}'|$ can then be written as $\vec{p}/|\vec{p}|=(\sin\omega, 0, \cos\omega)$ and $\vec{p}'/|\vec{p}'|=(\sin\theta'\cos\phi, \sin\theta'\sin\phi, \cos\theta')$, where $\omega$ and $\theta'$ are the polar angle of $\vec{p}$ and $\vec{p}'$, respectively, and $\phi$ is the azimuthal angle of $\vec{p}'$. The angle between $\vec{p}$ and $\vec{p}'$, denoted by $\theta$, can thus be expressed in terms of $\omega$ and $\theta'$:
\begin{eqnarray}
  \label{eq:angles}
  \cos\theta = \sin\theta' \sin\omega \cos\phi + \cos\theta' \cos\omega.
\end{eqnarray}
We can now expand and rewrite the $\sum_{\vec{p}_{i'}} f_{i,i'}(p_{F_i}, \vec{p}_{i'}) \delta \bar{n}_{i'}$ term as
\begin{samepage}
\begin{eqnarray}
  \label{eq:interactions}
  \sum_{\vec{p}_{i'}} f_{i,i'}(p_{F_i}, \vec{p}_{i'}) \delta \bar{n}_{i'} & = & 
  \frac{g}{(2\pi)^3} \int dp' d\theta' d\phi p'^2 \sin\theta' f_{i,i'}(p_{F_i}, \vec{p}_{i'}) \delta \bar{n}_{i'} \nonumber \\
  & = & \frac{g}{(2\pi)^3} \left(\int_{\cos\theta'>0} d\theta' d\phi p_{F_{i'}}^2 \sin\theta' f_{i,i'}(p_{F_i}, p_{F_{i'}}) \delta p_{i'} - \int_{\cos\theta'<0} d\theta' d\phi p_{F_{i'}}^2 \sin\theta' f_{i,i'}(p_{F_i}, p_{F_{i'}}) \delta p_{i'} \right) \nonumber \\
  & = & \frac{g}{(2\pi)^3} \mu_{i'} u\left(\int_{0}^{\pi/2} d\theta' d\phi p_{F_{i'}}^2 \sin\theta'\cos\theta' f_{i,i'}(p_{F_i}, p_{F_{i'}}) + \int_{\pi/2}^{\pi} d\theta' d\phi p_{F_{i'}}^2 \sin\theta'\cos\theta' f_{i,i'}(p_{F_i}, p_{F_{i'}}) \right) \nonumber \\
  & = & \frac{g}{(2\pi)^3} \mu_{i'} u p_{F_{i'}}^2 \int_{0}^{\pi} d\theta' d\phi \sin\theta'\cos\theta' f_{i,i'}(p_{F_i}, p_{F_{i'}}) \nonumber \\
  & = & \frac{g}{(2\pi)^3} \mu_{i'} u p_{F_{i'}}^2 \int_{0}^{\pi} d\theta' d\phi \sin\theta'\cos\theta' (f_{0,ii'} + f_{1,ii'}\cos\theta) \nonumber \\
  & = & \frac{g}{(2\pi)^3} \mu_{i'} u p_{F_{i'}}^2 f_{1,ii'} \int_{0}^{\pi} d\theta' d\phi \sin\theta'\cos\theta' (\sin\theta' \sin\omega \cos\phi + \cos\theta' \cos\omega) \nonumber \\
  & = & \frac{g}{4\pi^2} \mu_{i'} u p_{F_{i'}}^2 f_{1,ii'} \int_{0}^{\pi} d\theta' \sin\theta'\cos^2\theta' \cos\omega \nonumber \\
  & = & \frac{g}{4\pi^2} \frac{2}{3} \mu_{i'} u p_{F_{i'}}^2 f_{1,ii'} \cos\omega.
\end{eqnarray}
\end{samepage}
Plugging this result into Eq.~(\ref{eq:lb_fl2}), we obtain
\begin{eqnarray}
  \label{eq:emass}
  p_{F_i} & = & \sum_{i'} \frac{g}{4\pi^2} \frac{2}{3} \mu_{i'} p_{F_{i'}}^2 f_{1,ii'} + \mu_i \frac{de_i}{dp_i} \biggr|_{p_{F_i}} = \sum_{i'} \frac{g}{6\pi^2} \mu_{i'} p_{F_{i'}}^2 f_{1,ii'} + \mu_i \frac{p_{F_i}}{m^\ast_i},
\end{eqnarray}
and
\begin{eqnarray}
  \label{eq:emass2}
  m^\ast_i & = & \mu_i + \sum_{i'} g \frac{f_{1,ii'}(p_{F_i}, p_{F_{i'}})}{6\pi^2} \frac{\mu_{i'} p_{F_{i'}}^2 m_i^\ast}{p_{F_i}}.
\end{eqnarray}
\twocolumngrid

\bibliographystyle{apsrev4-1}
\bibliography{FLT.bib}

\end{document}